\documentclass[%
reprint,
aps,
]{revtex4-1}
\usepackage{graphicx}
\usepackage{hyperref}
\usepackage{color}
\usepackage{multirow}
\usepackage{dcolumn}
\usepackage{bm}
\usepackage{amsmath}
\usepackage{natbib}
\usepackage{docmute}
\usepackage{amssymb}
\usepackage{ulem}

\graphicspath{{./graph/}}

\usepackage{bm}
\newcommand{\diff}{d^2}

\newcommand{\imu}{{\rm i}}

\newcommand{\sgn}{\mathrm{sgn}\,}

\newcommand{\dg}{\dagger}
\newcommand{\la}{\langle}
\newcommand{\ra}{\rangle}
\newcommand{\al}{\alpha}

\begin{document}

\title{
Odd-Frequency Pairs in Chiral Symmetric Systems:\\
Spectral Bulk-Boundary Correspondence and Topological Criticality
}

\author{Shun Tamura$^{1}$, Shintaro Hoshino$^{2}$ and Yukio Tanaka$^{1}$}
\affiliation{%
$^1$Department of Applied Physics, Nagoya University, Nagoya 464-8603, Japan\\
$^2$Department of Physics, Saitama University, Saitama 338-8570, Japan
}

\date{\today}

\begin{abstract}


   Odd-frequency Cooper pairs with chiral symmetry
   emerging at the edges of topological superconductors are 
   a useful physical quantity for characterizing the topological properties
   of these materials. In this work, we show
   that the odd-frequency Cooper pair amplitudes can be expressed 
   by a winding number extended to a nonzero
   frequency, which is called a 
   ``spectral bulk-boundary correspondence,'' 
   and can be evaluated from the spectral features of the bulk.
   The odd-frequency Cooper pair amplitudes are
   classified into two categories: 
   the amplitudes in the first category have the singular functional form 
   $\sim 1/z$ (where $z$ is a complex frequency) that reflects
   the presence of a topological surface Andreev bound state, whereas
   the amplitudes in the second category have the regular form $\sim z$ 
   and are regarded as
   non-topological.
   We discuss the topological phase transition by using the coefficient
in the latter 
   category, which undergoes a power-law divergence
   at the topological phase transition point
and
   is used to indicate
   the distance to the critical point.
   These concepts are established based on 
   several concrete models, including a Rashba nanowire 
   system that is promising for 
   realizing Majorana fermions.

\end{abstract}
\pacs{pacs}

\maketitle
\thispagestyle{empty}

{\it Introduction}.---
The findings of quantum Hall systems and topological insulators have introduced 
topology into condensed matter physics~\cite{PG87,Volovikbook,Hasan},
leading to the discovery of a 
host of topological materials.
One 
important property
of topological systems is that the number of 
edge modes including 
the zero energy state is predicted by the topological number, which is
defined 
by the bulk~\cite{RH02,ryu10,qi11,SatoFujimoto2016,RevModPhys.88.035005,SatoAndo2017,PhysRevB.97.115143}.
This relation is called the
``bulk-boundary correspondence'' and has been a key concept in 
condensed matter physics~\cite{Hatsugai93,SRFL08}. 

The surface Andreev bound states (SABSs)
in topological superconductors
are associated with 
a nontrivial topological number,
and some 
are 
Majorana fermions~\cite{STYY11,tanaka12}. 
In terms of Cooper pairs, the SABSs 
indicate
the presence of 
odd-frequency Cooper pairs 
at the boundary, which have an odd functional 
form in time
and frequency~\cite{odd3,odd3b,tanaka12}.
Such exotic Cooper pairing was
first proposed by Berezinskii~\cite{Berezinskii}, and the corresponding
realization was discussed not only 
in the bulk state~\cite{Belitz1,Balatsky,Emery,Coleman} 
but also in a number of systems such as superconducting junctions 
based on
ferromagnets~\cite{Efetov1,Eschrig2003,Eschrig2015}, 
diffusive normal metals~\cite{odd1}, 
and 
vortex cores~\cite{Yokoyamavortex,Tanuma09}.
In addition, their peculiar paramagnetic responses have also been 
discussed~\cite{Meissner10,SuzukiAsano1,SuzukiAsano2,Lutchyn2017,Bernardo2}, and
the odd-frequency pairing 
has become a 
topic of interest in condensed matter physics~\cite{tanaka12,Linder17,PhysRevB.97.134523}.
For 
SABSs in topological superconductors, 
the relevance of the odd-frequency pairing 
is known~\cite{Asano2013,Stanev2014,Sau2015,Klinovaja2016,Crepin,Cayao,Keidel,PhysRevB.97.134523,Cayao2018};
the pair amplitude has a 
\textit{singular} functional form and diverges at 
zero frequency, $F^{\rm odd}_{\rm edge}(z) \sim 1/z$,
with complex frequency $z$.
This formula is
distinct from the \textit{regular} 
form
[$F^{\rm odd}_{\rm edge}(z) \sim z $], which appears ubiquitously 
because of the broken symmetry 
(e.g., the absence of translational symmetry at the edge)~\cite{odd3,odd3b,Eschrig2007}.
Thus, the topological superconducting systems offer a unique 
testing ground to develop ways to control the properties 
of odd-frequency Cooper pairing and 
to improve our understanding of Cooper pairs.

The search for
topological superconductivity has led to intensive studies of
the chiral symmetric systems~\cite{STYY11,TMYYS10,YSTY10,Volovik2011,Brydon11,Schnyder2012,Tewari2012L,Wong}, 
including
Rashba nanowire systems, which are promising for 
experimental realization of 
Majorana fermions at the edge~\cite{lutchyn10,oreg10}
and expected for a platform of a topological quantum computing~\cite{Kitaev01,BRAVYI2002210,Sarma2015}.
The chiral operator $\Gamma$
anticommutes with the Hamiltonian 
($\{\Gamma,{\cal H}\}=0$).
The index theorem tells us that the winding number defined 
in the bulk predicts the number of SABSs 
via the bulk-boundary correspondence~\cite{STYY11}.
The chiral symmetric systems also include a non-superconducting topological insulator such as a
Su-Schrieffer-Heeger 
(SSH) model~\cite{PhysRevLett.42.1698,RevModPhys.60.781}
and Shockley model~\cite{PhysRev.56.317,PhysRevB.86.075304}.

In this Letter, 
we extend
the 
bulk-boundary correspondence 
from zero frequency to nonzero frequency, 
which we call ``spectral bulk-boundary correspondence'' (SBBC).
By using the SBBC, 
the odd-frequency Cooper pairs accumulated at the boundary can be evaluated 
from the physical quantity 
determined in the bulk over the entire frequency range.
We further clarify that the regular odd-frequency Cooper pair 
amplitude ($\sim z $) can be used as {\it a degree of proximity to the topological phase transition}, 
which is analogous to 
using the susceptibility 
as a degree of the proximity 
to the phase transition in standard statistical physics.
The coefficient follows a power-law divergence at the topological phase transition, which reveals
the topological criticality.
Thus, we identify the fluctuation behavior associated with the topological transition, 
and can even go beyond the simple integer classification of phases.

\begin{figure}[t]
   \centering
   \includegraphics[width = 8.5cm]{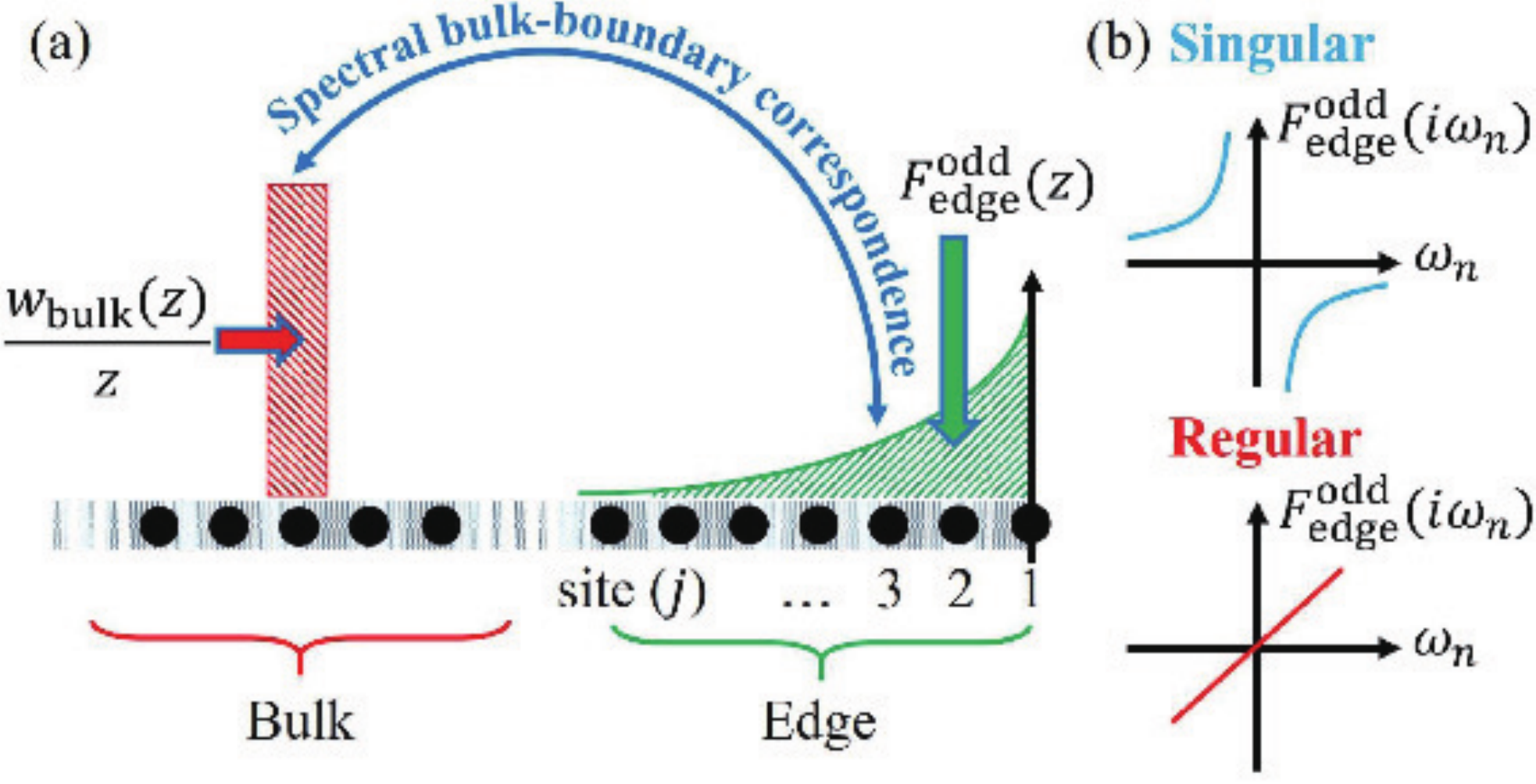}
   \caption{%
      (a) Schematic of the correspondence between 
      odd-frequency Cooper pair correlation 
      $F_\mathrm{edge}^\mathrm{odd}(z)$ and the extended winding number $w_\mathrm{bulk}(z)$ 
      with complex frequency $z$.
      (b) Graphs of singular and regular 
      frequency dependence of $F_\mathrm{edge}^\mathrm{odd}$.
      $z=i\omega_n$ is a Matsubara frequency and if $z$ is purely imaginary, $F_\mathrm{edge}^\mathrm{odd}$ 
      is also purely imaginary.
   }%
   \label{fig:schematic}
\end{figure}

{\it SBBC}.---
We begin by demonstrating the SBBC\@. 
The following relation holds for chiral symmetric systems and for
any complex frequency $z \in \mathbb C$:
\begin{align}
   F_\mathrm{edge}^\mathrm{odd}(z)
   =&
   \frac{w_\mathrm{bulk}(z)}{z},
   \label{eq:BEC}
\end{align}
with
\begin{align}
   F_\mathrm{edge}^\mathrm{odd}(z)
   =&
   \underset{j}{\mathrm{Tr}}
   \left[
      \Gamma G(z)
   \right],
   \label{eq:LH_k}
   \\
   w_\mathrm{bulk}(z)
   =&
\tfrac{\imu}{2}
   \underset{k}{\mathrm{Tr}}
   \left[
      \Gamma G(z)\partial_k
G^{-1}(z)
   \right]
   \label{eq:RH_k}
\end{align}
where 
the trace
in Eq.~(\ref{eq:LH_k}) is taken 
over a semi-infinite space:
${\rm Tr}_j \cdots = {\rm tr}\sum_{j=1}^{\infty} \la j | \cdots |j\ra$.
The surface is located 
on the right side, 
as shown in Fig.~\ref{fig:schematic}(a), and the site index 
is a positive integer.
The Green's function is defined by $G(z)=1/(z-{\cal H})$ with the Hamiltonian $\mathcal H$.
The trace ${\rm tr}$ is taken over the internal degrees of freedom 
composed of, e.g., spin and orbital indices.
On the other hand,
the trace $\mathrm{Tr}_k$ in Eq.~(\ref{eq:RH_k}) 
is taken over the bulk labeled by the wave vectors: 
${\rm Tr}_k \cdots = {\rm tr} \int_{-\pi}^{\pi} \frac{dk}{2\pi} \la k | \cdots | k\ra$.
A schematic 
of $F_\mathrm{edge}^\mathrm{odd}(z)$ and 
$w_\mathrm{bulk}(z)$ appears in Fig.~\ref{fig:schematic}(a).

In the zero-frequency limit, 
$w_{\rm bulk}(z\rightarrow 0)\equiv W$ 
is identified as the winding number~\cite{PhysRevB.83.085426}.
The full profile of $w_{\rm bulk}(z)$ is then regarded as an extension of the winding number to nonzero frequency.
The quantity $F^{\rm odd}_{\rm edge} (z)$ is the off-diagonal component 
of the Green's function located at the edge and represents the pair amplitude for topological superconductors.
We confirm that both sides of Eq.~(\ref{eq:BEC}) are odd in $z$,
meaning that $F^{\rm odd}_{\rm edge} (z)$ relevant to the SBBC is
an odd-frequency Cooper pair amplitude.
At zero frequency, Eq.~\eqref{eq:BEC} connects the nontrivial topological number $W\neq 0$ 
to
the odd-frequency pair through the
singular functional form $F^{\rm odd}_{\rm edge} (z\rightarrow 0) = W/z$ [Fig.~\ref{fig:schematic}(b)].
Furthermore, Eq.~\eqref{eq:BEC} shows that this connection persists to finite frequencies, 
which extends the concept of 
conventional frequency-independent
bulk-boundary correspondence~\cite{PhysRevB.48.11851}.
This means that the total amount of the odd frequency Cooper pair correlation accumulated
near the surface is predicted by the bulk value [$w_\mathrm{bulk}(z)/z$].

\begin{figure}[t]
   \centering
   \includegraphics[width = 7.5cm]
   {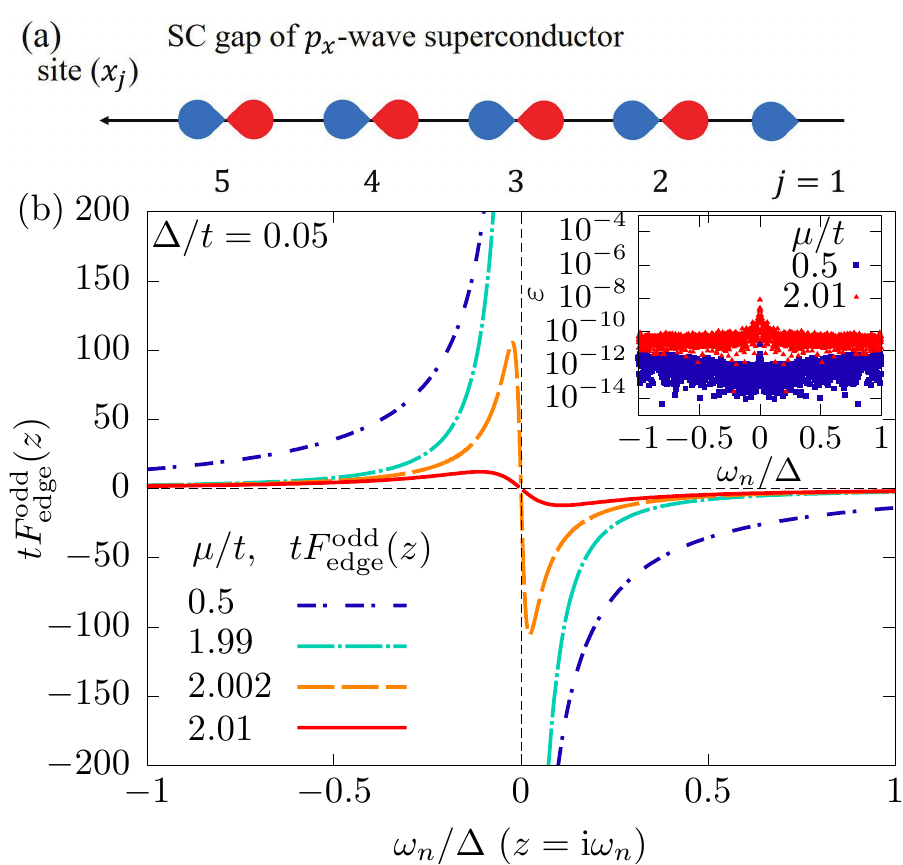}
   \caption{%
      (a) Schematic of one-dimensional $p$-wave pairing Kitaev chain. 
      (b) The imaginary part of total amount of the odd frequency Cooper pair amplitude 
      $\mathrm{Im}[F_\mathrm{edge}^\mathrm{odd}(z)]$ is plotted as a function
      of $\omega_n=z/i$ at $\Delta/t=0.05$ with several $\mu$.
      In the inset, $\varepsilon$ is plotted as a function of 
      $\omega_n$ for $\mu/t=0.5$ and 2.01.
   }%
   \label{fig:pic_kitaev_odd_freq}
\end{figure}

Equation~\eqref{eq:BEC} is confirmed exactly in 
various limits of a Kitaev chain~\cite{Kitaev01}, 
and is also confirmed numerically for various
chiral symmetric systems such as Rashba nanowires~\cite{lutchyn10,oreg10}, and two-dimensional 
$d_{x^2-y^2}$-wave superconductors~\cite{YSTY10,kashiwaya00}.
We first take a closer look at the SBBC in the Kitaev chain
which is a one-dimensional $p$-wave superconductor with fully polarized spins 
[see Fig.~\ref{fig:pic_kitaev_odd_freq}(a)].
The Hamiltonian is ${\cal H}=\sum_k C_k^\dag H(k)C_k$ with
\begin{align}
   H(k)
   =&
   -(t\cos k +\mu/2)
   \tau_3
   -
   \Delta
   \sin k
   \tau_2
   \label{eq:Kitaev_chain}
\end{align}
and $C_k={(c_k,c_{-k}^\dag)}^\mathrm{T}$ where $c_k$ is an annihilation operator of electrons.
$t$ is a hopping integral and $\mu$ is a chemical potential,
$\tau_\mu$ ($\mu=1,2,3$) is a Pauli
matrix in Nambu space, and $\Delta$ is the
$p$-wave superconducting gap.
The condition for a 
topological superconductor is 
$|\mu|<\mu_\mathrm{c}=2t$.
We now show that $F_\mathrm{edge}^\mathrm{odd}(z)$ 
corresponds to the odd-frequency Cooper pair amplitude.
We first set $z=i\omega_n$, where $\omega_n$ is an imaginary (Matsubara) frequency.
The chiral operator $\Gamma$ for a semi-infinite system is
$   \Gamma
   =
   \mathrm{diag}
   (\ldots,\tau_1,\tau_1),
$
where $\tau_1$ is a Pauli matrix acting on a Nambu space $(c_j,\  c_j^\dg)$.
The function $F_{\rm edge}^{\rm odd}(z)$ is then
\begin{align}
F_{\rm edge}^{\rm odd}(\imu \omega_n)
   =&
   \int_0^\mathrm{\beta} \mathrm{d}\tau
   e^{\mathrm{i}\omega_n\tau}
   \sum_{j=1}^{\infty}
\langle
   c_j(\tau)c_j(0) \rangle
+ {\rm c.c.},
\end{align}
where the right-hand side is the Cooper pair amplitude for $s$-wave spin-triplet superconductivity 
and must be an odd function in time and frequency 
to satisfy the Pauli exclusion principle.

When the special conditions $t=\Delta$ and $\mu=0$ are satisfied, Majorana fermions are localized at the edges of the system with zero localization length.
In this case, the SBBC relation (\ref{eq:BEC})
takes the analytical form
\begin{align}
   F_\mathrm{edge}^\mathrm{odd}(z)
   =&
   \frac{w_\mathrm{bulk}(z)}{z}
   =
   \frac{-t^2}{z(z^2-t^2)}.
\end{align}
See the supplementary material (SM) 
for more a detailed derivation [SM from I-A to I-E].
We can also construct the quasi-classical Green's function 
for a coherence length sufficiently large
compared with the inverse 
Fermi momentum.
The SBBC also takes the following analytical form
in this limit [SM I-F]:
\begin{align}
   F_\mathrm{edge}^\mathrm{odd}(z)
   =&
   \frac{w_\mathrm{bulk}(z)}{z}
   =
   \frac{\Delta}{z\sqrt{\Delta^2-z^2}}.
   \label{eq:BEC_continuous}
\end{align}
In Eq.~(\ref{eq:BEC_continuous}), 
we assume $\mu\gg |z|$ and $\mu\gg\Delta$ 
where we measure the chemical potential from the bottom of the band.
Let us also consider the numerical results for the Kitaev chain with $\Delta/t=0.05$ and $\mu\neq0$.
Figure~\ref{fig:pic_kitaev_odd_freq}(b) plots
$F_\mathrm{edge}^\mathrm{odd}(z)$
as a function of 
$\omega_n$
for several $\mu/t$.
The parameters $\mu/t=0.5$ and $1.998$ are located in 
the topological region: 
in the limit 
$\omega_n\rightarrow 0$, it diverges as 
$\omega_n$ approaches zero (singular).
For $\mu/t=2.002$ and 2.01, however, 
it approaches zero for $\omega_n\rightarrow0$ (regular).
To check the numerical accuracy of the SBBC we calculate the 
quantity
$  
\varepsilon (z)
=
|{F}_{\mathrm{edge}}^\mathrm{odd}(z)-{w}_{\mathrm{bulk}}(z)/z|/
{|{w}_{\mathrm{bulk}}(z)/z|}
$, which
is less than $10^{-10}$ for $\mu/t=0.5$ and 1.998 
(i.e., the same within numerical error).
For $\mu/t=2.002$ and 2.01, 
$\varepsilon<10^{-8}$ [SM I-G].
We also checked the SBBC for spatially changing
pair potentials near the edge, which are discussed in detail in the SM I-H.\

\begin{figure}[t]
   \centering
   \includegraphics[width = 8.5cm]{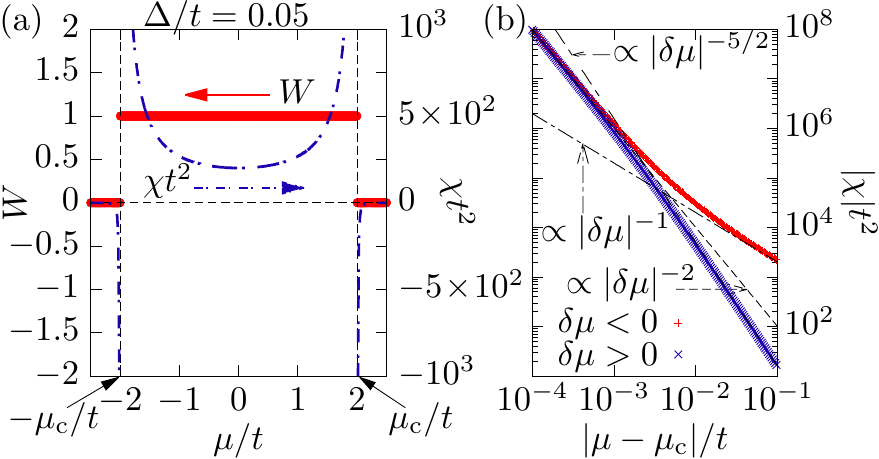}
   \caption{%
      (a) The winding number $W$ (left vertical axis) and $\chi$ (right vertical axis)
      are plotted as a function of $\mu$ for $\Delta/t=0.05$.
      (b) $|\chi|$ is plotted near $\mu_\mathrm{c}$ as a function of 
      $|\mu-\mu_\mathrm{c}|/t$ for $\mu>\mu_\mathrm{c}$ and $\mu<\mu_\mathrm{c}$
      with $\mu_\mathrm{c}=2t$ and $\delta\mu=\mu-\mu_\mathrm{c}$.
   }%
   \label{fig:pic_kitaev_odd_freq2}
\end{figure}

{\it Topological criticality in odd-frequency Cooper pairs}.---
%
The topological criticality has been discussed 
in terms of the physical quantities such as divergent correlation length, 
compressibility~\cite{Mondragon-Shem14,Altland14,PhysRevA.87.063618,PhysRevA.87.063618,PhysRevB.92.104514,PhysRevB.85.155302, PhysRevB.98.035419}.
Here we demonstrate that the criticality appears also in the odd-frequency Cooper pairs.
In the low-frequency limit, the odd frequency Cooper pair amplitude is
\begin{align}
   F_{\rm edge}^{\rm odd} (z) &= \frac{W}{z} + \chi z + O(z^3).
   \label{eq:expand_F}
\end{align}
The first term on the right-hand side 
represents the singular odd-frequency pair, and the second term
represents the regular pair.
The quantity $W = w_{\rm bulk}(z=0)$ is a standard winding number 
defined in the bulk.
By using the SBBC, $\chi$ can be expressed as a bulk quantity:
$
   \chi = \tfrac{\imu}{2} 
{\rm Tr}_k \left[ \Gamma H^{-3} \partial_k H  \right]
$.
We find that at the topological quantum phase transition where $W$ changes,
the coefficient $\chi$ undergoes a 
power law divergence upon approaching 
from either side of the phases.
As shown in Fig.~\ref{fig:pic_kitaev_odd_freq2}(b), the critical behavior 
in the limit $\mu\to \mu_\mathrm{c}$ is $ \chi \sim |\mu-\mu_\mathrm{c}|^{-2}$.
However, this exponent crossovers into another ones, namely $-1$ for $\mu<\mu_\mathrm{c}$ 
and $-5/2$ for $\mu>\mu_\mathrm{c}$ [SM I-I], and thus the behavior is quite asymmetric around 
the critical point $\mu=\mu_\mathrm{c}$ as in Fig.~\ref{fig:pic_kitaev_odd_freq2}(a).
While the exponent $-2$ is consistent with the Ising universality~\cite{Sachdev_2011}, for the exponents $-1$ and $-5/2$, one needs a generalization of the concept.

In order to understand the above critical behaviors, 
we use scaling theory for the effective action.
The effective low-energy action 
is introduced as
\begin{align}
   \hspace{-1.5mm}S\hspace{-0.5mm}
   =\hspace{-1mm}
   \int\hspace{-1mm} \mathrm{d}\tau \mathrm{d}x\hspace{-0.5mm}
   \Big[
      \psi^\dag \partial_\tau \psi
      \hspace{-0.5mm}
      +
      \hspace{-0.5mm}
\frac{v}{2}
      \left(
         \psi^\dag \partial_x \psi^\dag
         \hspace{-0.5mm}
         +
         \hspace{-0.5mm}
         {\rm H.c.}
      \right)
      \hspace{-0.5mm}
      +
      \hspace{-0.5mm}
      m
      \psi^\dag \psi
      \hspace{-0.5mm}
      - 
      \hspace{-0.5mm}
      \Lambda \psi^\dg \partial_x^2 \psi
   \Big],
\label{eq:action}
\end{align}
where $v$ is a velocity, $m$ is a mass and $\Lambda$ is a coefficient of the 
second derivative term.
For the Kitaev chain, $v$, $m$ and $\Lambda$ are given by
$v=\Delta
$, $m=\delta \mu=\mu-\mu_\mathrm{c}$ and $\Lambda=t$, respectively.
The corresponding energy is given by 
$\lambda_k^2  = {(m+ 
\Lambda k^2)}^2 + 
v^2 k^2$.
Usually the term with $\Lambda$ is irrelevant and can be neglected and one obtains 
the Ising universality.
However, in superconductors $v=\Delta$ is an energy
gap and hence not only $m$ 
(distance to critical point) but also $v$ are much smaller than $\Lambda$.
In this case, the quadratic term must be kept, to result in a variety of 
critical behaviors (see Fig.~\ref{fig:pic_kitaev_odd_freq2}).
We perform the scale transformation as
$x'=xe^{-l}$ ($\dim [x]=-1$).
The action in the low-temperature limit is invariant if the scaling dimensions satisfy
$\dim [\psi]=1/2$, $\dim[\tau]=-2$ (dynamical critical exponent), $\dim[v]=1$, and $\dim[m]=2$.

\begin{figure}[t]
   \centering
   \includegraphics[width = 7.6cm]{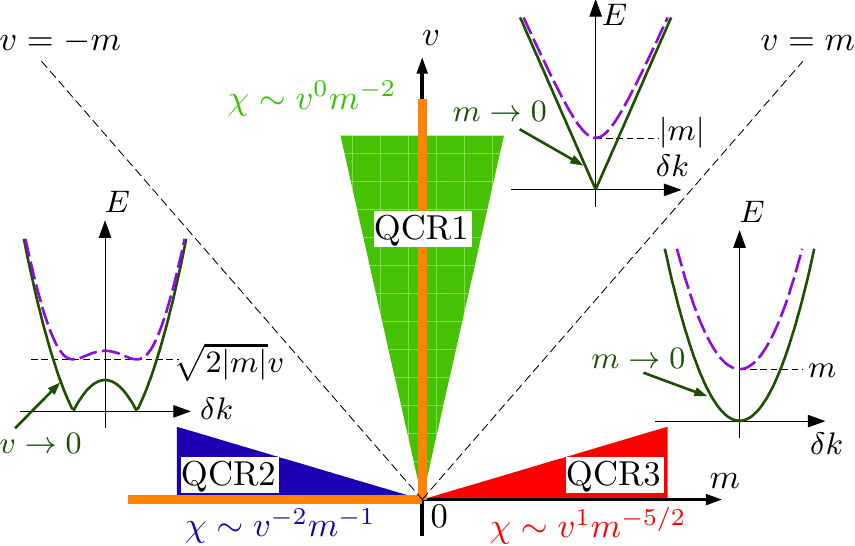}
   \caption{%
Critical regimes for the Kitaev chain near $\mu=\mu_\mathrm{c}$
      and $\Delta=0$, illustrated in the plane of the mass $m(\propto \mu-\mu_{c})$ and velocity $v(\propto\Delta)$.
      Colored regions 
         (QCR1 to QCR3)
      are characterized by different critical exponents $\chi \sim v^\al m^\beta$ with fixed $\al + 2\beta = -4$.
      The orange line shows the critical points at which the energy gap closes.
   }\label{fig:phase_diag_crit}
\end{figure}

Now let us consider the generalized winding number in the expansion form 
$w_\mathrm{bulk}(z) = \sum_{n=0}^\infty a_{2n} z^{2n}$ (specifically, $a_0=W$ and $a_2=\chi$).
In the critical region
$|\Lambda|\gg |m|, |v|$, we can express the coefficients as
$a_{2n}(m, v) = a'_{2n} v^{\al_n} m^{\beta_n}$.
Using the fact that the scaling dimension of $w_\mathrm{bulk}(z)$ is zero, 
we have the relation between $\al_n$ and $\beta_n$:
\begin{align}
   \al_n \dim [v] + \beta_n \dim [m] + 2 n \dim [z] = 0.
   \label{eq:relation}
\end{align}
The odd-frequency pair amplitude can now be written as 
\begin{align}
   F^\mathrm{odd}_\mathrm{edge}(z) 
   = 
   \frac{1}{z} \mathcal W 
   \left( 
      \frac{v}{\sqrt{\Lambda m}}, \frac{z^2}{m^2} 
   \right),
\end{align}
by using the SBBC.\
Namely, the odd-frequency pair amplitude is generally a function of three independent 
variables ($z,m,v$ with $\Lambda$ being unit of energy), but for $|m|,|v| \ll |\Lambda|$ 
they are reduced to two variables.
From the shapes of the energy spectrum shown in Fig.~\ref{fig:phase_diag_crit},
we can identify the three regimes,
in which $F^\mathrm{odd}_\mathrm{edge}(z) $ can be written by a single scaling 
function with only one variable.
Correspondingly we obtain three quantum critical regions (QCR) QCR1, QCR2
and QCR3 shown in Fig.~\ref{fig:phase_diag_crit} [detailed derivation is shown in SM I-J], which gives exponents
$\chi$ behaving as $v^0m^{-2}$ in QCR1, $v^{-2}m^{-1}$ in QCR2 and $v^1m^{-5/2}$ in QCR3.
These critical exponents can also seen in Fig.~\ref{fig:pic_kitaev_odd_freq2}(b) 
Our results thus extend the conventional knowledge about topological phase transitions.
Since the low-energy effective actions for Rashba nanowire and $d$-wave superconductors 
are given by the same action as that in the Kitaev chain, we get the similar critical behaviors as 
demonstrated in the following.

\begin{figure}[t]
   \centering
   \includegraphics[width = 7.5cm]{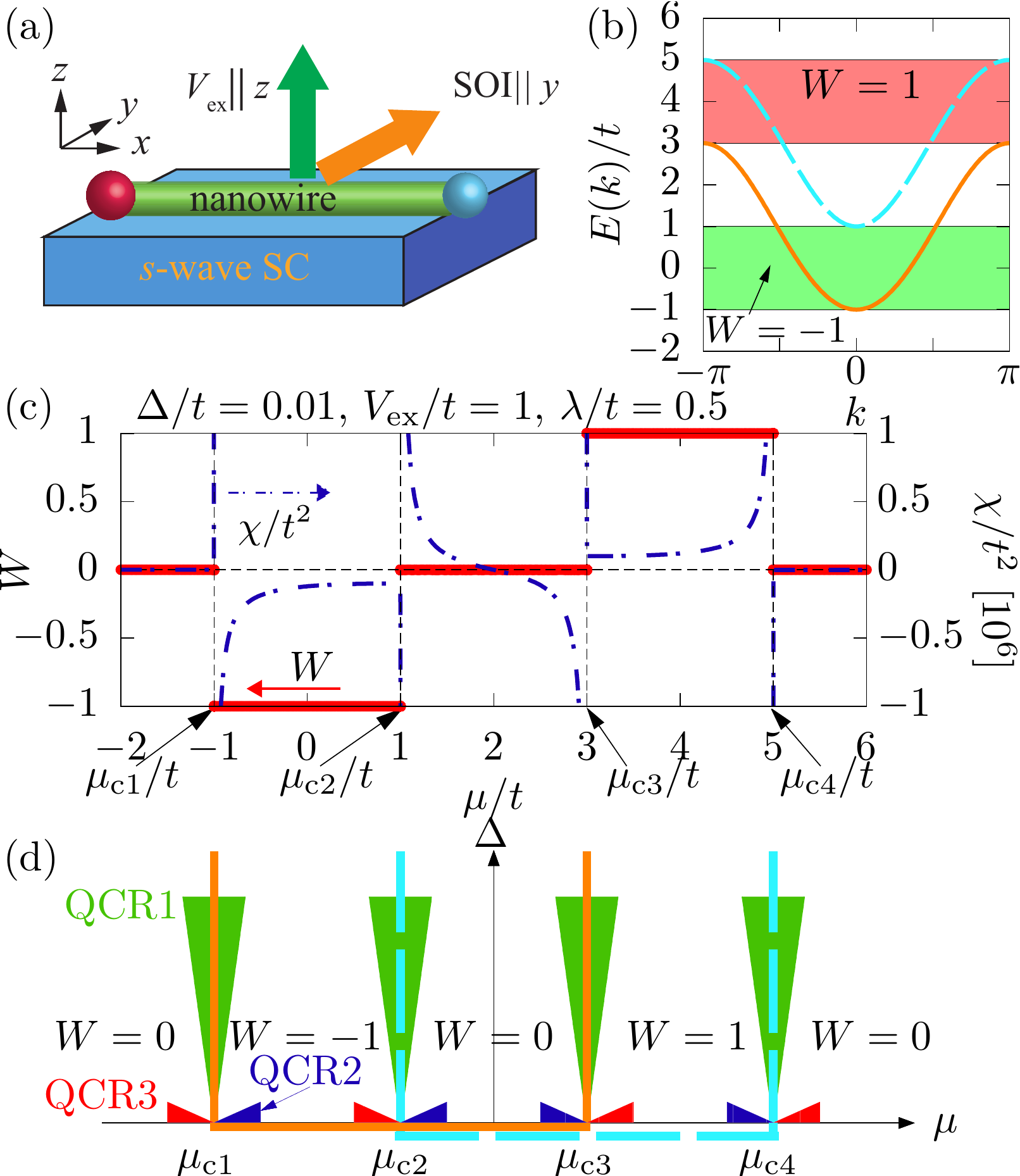}
   \caption{%
      (a) Schematic of Rashba nanowire.
      (b) Energy dispersion of nanowire 
      for $\Delta/t=0.01$, $V_\mathrm{ex}/t=1$, $\lambda/t=0.5$, and $\mu=0$.
      The red and green shaded areas are the topological regime.
      (c) $W$ (red, left vertical axis) and $\chi$ (blue, right vertical axis) 
      are plotted as a function of $\mu/t$. 
      $\mu_{c1}=-\sqrt{V_\mathrm{ex}^2-\Delta^2}$,
      $\mu_{c2}=\sqrt{V_\mathrm{ex}^2-\Delta^2}$,
      $\mu_{c3}=4t-\sqrt{V_\mathrm{ex}^2-\Delta^2}$,
      and
      $\mu_{c4}=4t+\sqrt{V_\mathrm{ex}^2-\Delta^2}$.
      (d) Schematic of the QCRs.\
   }%
   \label{fig:pic_nanowire}
\end{figure}

{\it Rashba nanowire}.---
The SBBC and singular behavior of the regular odd-frequency pair amplitude
can also be seen in the other models; e.g., 
the one-dimensional Rashba nanowire on an
$s$-wave superconductor~\cite{lutchyn10,oreg10}, where a Majorana fermion located at the edge
is accompanied by odd-frequency pairing~\cite{Asano2013,Stanev2014} [see Fig.~\ref{fig:pic_nanowire}(a)].
The Hamiltonian is given by $   {\cal H}
   =
   \frac{1}{2}
   \sum_k
   C_k^\dag H(k) C_k,
$
with
\begin{align}
   H(k)
   =&
   \left[
      \varepsilon(k)\sigma_0
      +
      V_\mathrm{ex}\sigma_3
      +
      \lambda\sin k\sigma_2
   \right]
   \tau_3
   +
   \Delta \mathrm{i}\sigma_2\mathrm{i}\tau_2,
\end{align}
where 
$   C_k
   =
   {
      \left(
         c_{k,\uparrow},c_{k,\downarrow},c_{-k,\uparrow}^\dag,c_{-k,\downarrow}^\dag
      \right)
   }^{\rm T}
$
,
$\varepsilon(k)=-2t\cos(k)+2t-\mu$, $V_\mathrm{ex}$ is a magnetic field, 
$\lambda$ is the Rashba spin orbit interaction,
and $\Delta$ is an $s$-wave superconducting gap.
$\sigma_\mu$ ($\mu=1,2,3$) is a Pauli matrix in spin space.
The system is located in the topological regime when $\sqrt{\mu^2+\Delta^2}<|V_\mathrm{ex}|<\sqrt{{(4t-\mu)}^2+\Delta^2}$.
The chiral operator can be defined provided
the magnetic field and the spin-orbit interaction are orthogonal:
$\Gamma=\sigma_0\tau_1$.
The energy dispersion of the Rashba nanowire is shown in 
Fig.~\ref{fig:pic_nanowire}(b).


We discuss criticality for Rashba nanowire.
Figure~\ref{fig:pic_nanowire}(c) shows $W$ and $\chi$ as a function of $\mu$.
The parameter $\chi$ diverges near the quantum transition points, 
showing topological criticality.
A very sharp divergence appears
when e.g., $\mu \rightarrow -\sqrt{V_\mathrm{ex}^2-\Delta^2} -0$, which is 
due to the small magnitude of the
superconducting gap (see also the SM II-A and II-B).

The criticality of the Rashba nanowire is understood by using 
the results for the Kitaev chain as a building block.
The energy dispersion in Fig.~\ref{fig:pic_nanowire}(b) is viewed
as two coupled nanowires;
there are two cosine-like dispersions
and they are regarded as double Kitaev chain.
$\mu_{c1}$ ($\mu_{c2}$) and $\mu_{c3}$ ($\mu_{c4}$) are the lower and upper
boundary of the energy dispersion shown by orange line (light blue dotted line).
The orange and light blue 
lines in Fig.~\ref{fig:pic_nanowire}(d) correspond  to the band edges at which energy gap closes 
(i.e., at critical point).
For small $\Delta$ ($\ll t$), which is usually satisfied in superconductors, 
the phase diagram in Fig.~\ref{fig:phase_diag_crit} can be applied for each band edge.
Then QCR2 with $\chi\sim {(\mu-\mu_\mathrm{c})}^{-1}$ [QCR3 with $\chi\sim {(\mu-\mu_\mathrm{c})}^{-5/2}$]
exists inside (outside) of each energy dispersion as shown in Fig.~\ref{fig:pic_nanowire}(d).
Thus complex and highly asymmetric behaviors of $\chi$ around topological 
transition are explained based on Fig.~\ref{fig:phase_diag_crit}.
Note that the sign of $\chi$ changes 
at $\mu/t=2$ in the non-topological regime.
This property reflects the situation in which
this non-topological phase is sandwiched 
between topological phases with the different winding numbers $W=+1$ and $W=-1$.
Namely, we can obtain information on the neighboring topological phases even 
in the non-topological phase by looking at 
the regular odd-frequency Cooper pairs.
In contrast, no such sign change appears for the Kitaev chain.

{\it $d$-wave superconductors}.---
We also calculate $W$ and $\chi$ for $d_{x^2-y^2}$-wave superconductor with
(11)-surface [shown in SM III-A].
In this case, $W$ changes its value as a function of a wave number which is 
parallel to the surface and $\chi$ also diverges at topological transition 
points [SM III-B]. 


{\it Conclusion}.---
We demonstrate that the SBBC $F_\mathrm{edge}^\mathrm{odd}(z)=w_\mathrm{bulk}(z)/z$
[Eq.~(\ref{eq:BEC})] holds for chiral symmetric systems such as the Kitaev chain,
Rashba nanowire
which is promising for the realization of the Majorana fermion,
and two dimensional $d_{x^2-y^2}$-wave superconductors.
The Cooper pair amplitude can be expanded to the form
$F_\mathrm{edge}^\mathrm{odd}(z)=W/z+\chi z+\mathcal{O}(z^3)$, where $W$ is a topological number.
We show that the coefficient $\chi$ diverges at the topological transition point and 
the critical behaviors are interpreted in terms of the effective action which generalizes the known Ising universality class.



{\it 
Note added}.---
After submission of our previous version [arXiv:1809.05687v1 (2018)], 
we are aware that Daido and Yanase have submitted a proof of the SBBC based on a 
chirality polarization [arXiv:1901.03482v1 (2019)].

\begin{acknowledgments}
{\it Acknowledgments}.---
   We are grateful to M.~Sato, S.~Kobayashi, T.~Imaeda and S.~Nakosai 
   for useful discussions.
   This work was supported by Grant-in-Aid
   for Scientific Research on Innovative Areas, Topological
   Material Science (Grants No. No.~JP15H05851, 
   No.~JP15H05853, and No.~JP15K21717) and Grant-in-Aid for
   Scientific Research B (Grant No. JP18H01176) 
   from the Ministry of Education, Culture,
   Sports, Science, and Technology, Japan (MEXT).
   This work was also supported 
   by Japan Society for Promotion of Science (JSPS) KAKENHI Grant No.~18K13490.
\end{acknowledgments}

\bibliography{TopologicalSC}

\pagebreak

\begin{widetext}
   \begin{center}
\textbf{\large
Supplemental Material for \\
``Odd-frequency pairs
in chiral symmetric systems:
spectral bulk-boundary correspondence and topological criticality''
}
   \end{center}
\end{widetext}

\setcounter{equation}{0}
\setcounter{figure}{0}
\setcounter{table}{0}
\setcounter{page}{1}
\makeatletter
\renewcommand{\theequation}{S\arabic{equation}}
\renewcommand{\thefigure}{S\arabic{figure}}
\section{Kitaev chain}
In~\ref{sec:Hamiltonian_kitaev}, a Hamiltonian for the Kitaev chain and 
a formal formula of $w_\mathrm{bulk}(z)$ are introduced.
Next four subsections are analytical results for $t=\Delta$.
In~\ref{sec:w_exact_kitaev}, an analytical formula for $w_\mathrm{bulk}(z)$ with $t=\Delta$ is given.
In~\ref{sec:surface_G_kitaev}, we explain the derivation of a surface Green's function 
for the Kitaev chain and
in~\ref{sec:sG_arbit_kitaev}, we give a local Green's function for arbitrary site.
In~\ref{sec:sG_mu0_kitaev}, the local Green's function for $\mu=0$ is given and
the spectral bulk-boundary correspondence (SBBC) is analytically shown.
In~\ref{sec:kitaev_cont}, we show the SBBC holds for the continuum model of the $p$-wave superconductor.
Following two subsections are numerical results for $F_\mathrm{edge}^\mathrm{odd}(z)$ 
and $w_\mathrm{bulk}(z)/z$.
In~\ref{sec:SBEC_mun0_kitaev}, we numerically calculate $F_\mathrm{edge}^\mathrm{odd}(z)$ 
and $w_\mathrm{bulk}(z)/z$ for $\Delta/t=1$ and $\Delta/t=0.05$, and
in~\ref{sec:SBEC_kitaev_sp_mod}, we introduce spatial modulation near the surface
of the gap function and the chemical potential.
We give an exact formula of $\chi$ for Kitaev chain with arbitrary $\Delta$ in~\ref{sec:chi_exact_kitaev}.
Finally, we explain the critical exponent of $\chi$ in~\ref{sec:chi_app}.
\subsection{Hamiltonians with open boundary and periodic boundary condition
\label{sec:Hamiltonian_kitaev}}
The Kitaev chain in the real-space basis is
\begin{align}
   {\cal H}
   =&
   -t\sum_j
   \left(
      c_j^\dag c_{j+1}
      +
      \mathrm{H.c.}
   \right)
   +
   \Delta
   \sum_j
   \left(
      c_{j+1}^\dag c_{j}^\dag
      +
      \mathrm{H.c.}
   \right)
   \nonumber
   \\
   &
   -
   \mu\sum_j c_j^\dag c_j,
   \label{eq:H_kitaev}
\end{align}
where $c_j$ is an annihilation operator on $j$-th site and $t$ is the hopping integral,
$\Delta$ is the $p$-wave superconducting gap and $\mu$ is the chemical potential.

For the open boundary (OB) condition with $n$-site system, the Hamiltonian is
\begin{align}
   {\cal H}^\mathrm{OB}_n
   =&
   C_n^\dag
   \begin{pmatrix}
      \hat{u} & \hat{t}
      \\
      \hat{t}^\dag & \hat{u} &\hat{t}
      \\
      &\hat{t}^\dag & \hat{u} &\ddots
      \\
      &&\ddots&\ddots&
   \end{pmatrix}
   C_n ,
   \\
   C_n
   =&
   \left(
      c_n,\:
      c_n^\dag,\:
      c_{n-1},\:
      c_{n-1}^\dag,\:
      \ldots,
      c_1,\:
      c_1^\dag
   \right)^\mathrm{T},
   \\
   \hat{u}
   =&
   \frac{1}{2}
   \begin{pmatrix}
      -\mu & 0
      \\
      0 & \mu
   \end{pmatrix},
   \\
   \hat{t}
   =&
   \frac{1}{2}
   \begin{pmatrix}
      -t & \Delta
      \\
      -\Delta & t
   \end{pmatrix}.
   \label{eq:def_hat_t}
\end{align}
We define a Green's function corresponding to ${\cal H}_n^\mathrm{OB}$ as
\begin{align}
   G_n(z)
   =&
   \frac{1}{z-{\cal H}_n^\mathrm{OB}}.
\end{align}

On the other hand, by imposing the periodic boundary condition on Eq.~(\ref{eq:H_kitaev}),
we obtain the Hamiltonian with the wave number basis as
\begin{align}
   {\cal H}_\mathrm{bulk}
   =&
   \sum_k C_k^\dag H_\mathrm{bulk}(k) C_k,
   \\
   C_k
   =&
   \left(
      c_{k}, \:c_{-k}^\dag
   \right)^\mathrm{T},
   \\
   H_\mathrm{bulk}(k)
   =&
   \left(
      -t\cos k-\frac{\mu}{2}
   \right)\tau_3
   -\Delta\sin k\tau_2,
\end{align}
with the Pauli matrices $\tau_\mu$ ($\mu=1,2,3$).
A Green's function with the periodic boundary condition is
\begin{align}
   g(z,k)
   =&
   \frac{%
      z
         +
      (-t\cos k-\frac{\mu}{2})\tau_3\
         -
      \Delta\sin k\tau_2
   }
   {%
      z^2-\Delta^2\sin^2 k-(-t\cos k-\frac{\mu}{2})^2
   }.
\end{align}

$w_\mathrm{bulk}(z)$ for the Kitaev chain is
\begin{align}
   w_\mathrm{bulk}(z)
   =&
   \frac{1}{2\pi}\int_{-\pi}^\pi \mathrm{d}k\:
   \frac{-t\Delta-\frac{\mu\Delta}{2}\cos k}
   {z^2-\Delta^2\sin^2 k-(-t\cos k-\frac{\mu}{2})^2},
   \label{eq:w_bulk_Kitaev_gen}
\end{align}
with
\begin{align}
   \Gamma  
   =& 
   \tau_1,
   \\
   \partial_k g^{-1}(z,k)
   =&
   -t\sin k\tau_3+\Delta\cos k\tau_2.
\end{align}

\subsection{Exact formula of $w_\mathrm{bulk}(z)$ for Kitaev chain
   with $t=\Delta$
\label{sec:w_exact_kitaev}}
We calculate the exact formula of $w_\mathrm{bulk}(z)$
for the Kitaev chain with $t=\Delta$ and an exact formula of $\chi$ can be obtained.
$w_\mathrm{bulk}(z)$ with $t=\Delta$ is given by
\begin{align}
   w_\mathrm{bulk}(z) 
   =&
   \frac{1}{2\pi}
   \int_{-\pi}^{\pi}
   \mathrm{d}k
   \frac{-t^2-\frac{1}{2}\mu t\cos k}
   {z^2-t^2-t\mu\cos k-\frac{1}{4}\mu^2}
   \nonumber\\
   =&
   \frac{1}{2}
   -
   \frac{-z^2-t^2+\frac{1}{4}\mu^2}
   {2\sqrt{(-z^2+t^2+\frac{1}{4}\mu^2)^2-(t\mu)^2}}.
   \label{eq:kitaev_w_exact}
\end{align}
Then $\chi$ is obtained by differentiate $w_\mathrm{bulk}(z)$ as
\begin{align}
   2\chi
   =&
   \left.
   \frac{\partial^2 w_\mathrm{bulk}(z)}{\partial z^2}
   \right|_{z=0}
   \nonumber\\
   =&
   \frac{1}{
      \left|
      -t^2
      +
      \frac{1}{4}\mu^2
      \right|
   }
   -
   \sgn( -t^2 + \frac{1}{4}\mu^2)
   \frac{t^2+\frac{1}{4}\mu^2}{
      \left|
      -t^2
      +
      \frac{1}{4}\mu^2
      \right|^2
   }
   \nonumber\\
   =&
   \frac{4\sgn(|\mu|-\mu_\mathrm{c})}{(\mu-2t)(\mu+2t)}
   \left[
      1
      -
      \frac{(4t^2+\mu^2)}
      {(\mu-2t)(\mu+2t)}
   \right],
   \label{eq:chi_Delta_t}
\end{align}
with $\mu_\mathrm{c}=2t$.
Then in the limit $\mu\rightarrow\mu_\mathrm{c}$, $\chi$ becomes
\begin{align}
   \lim_{\mu\rightarrow\mu_\mathrm{c}}|\chi|
   =&
   \frac{1}{|\mu-\mu_\mathrm{c}|^2}.
\end{align}

\subsection{Surface Green's function at rightmost site for $t=\Delta$\label{sec:surface_G_kitaev}}
In the following, we show the derivation of the surface Green's function for $t=\Delta$.
Note that for $t=\Delta$, $\det\hat{t}=0$ [$\hat{t}$ is given in Eq.~(\ref{eq:def_hat_t})] 
and the method to calculate the Green's function for 
the semi infinite system~\cite{Umerski} cannot be directly applicable.
For the calculation of the surface Green's functions with $t\neq\Delta$, 
we follow the procedure of Ref.~\onlinecite{Umerski}.

The recurrence relation of the local Green's function of the rightmost site 
for ${\cal H}_{n+1}^\mathrm{OB}$
[$G_{n+1}(z)$] and that for ${\cal H}_n^\mathrm{OB}$ [$G_{n}(z)$] is
\begin{align}
   &
   [G_{n+1}(z)]_{1,1}
   \nonumber\\
   =&
   \frac{1}{z-\hat{u}-\hat{t}^\dag [G_{n}(z)]_{1,1}\hat{t}}
   \nonumber\\
   =&
   \frac{1}{z^2-\frac{\mu^2}{4}-2zf^\mathrm{L}_n(z)}
   \begin{pmatrix}
      z-\frac{\mu}{2}-f^\mathrm{L}_{n}(z) & -f^\mathrm{L}_{n}(z)
      \\
      -f^\mathrm{L}_{n}(z) & z+\frac{\mu}{2}-f^\mathrm{L}_{n}(z)
   \end{pmatrix},
   \label{eq:rec_kitaev}
\end{align}
with 
\begin{align}
   f^\mathrm{L}_n(z)
   =
   \frac{t^2}{4}\sum_{l,m=1,2}\{[G_{n}(z)]_{1,1}\}_{lm}.
\end{align}
We use the same superscript L and R as in Ref.~\onlinecite{Umerski}.
Note that $[G_{n}(z)]_{1,1}$ is a $2\times2$ matrix 
and the index 1,1 indicates the rightmost site.
Then the recurrence relation for 
\begin{align}
   g^\mathrm{L}_n(z)
   \equiv
   2zf^\mathrm{L}_n(z)-t^2
   \label{eq:f_n}
\end{align}
is 
\begin{align}
   g^\mathrm{L}_{n+1}(z)
   =&
   \frac{t^2\mu^2/4}{z^2-\mu^2/4-t^2-g^\mathrm{L}_n(z)}.
   \label{eq:rec_kitaev2}
\end{align}
We obtain $g^\mathrm{L}_n(z)$ by using the M\"obius transformation as
\begin{align}
   g^\mathrm{L}_n(z)
   =&
   \begin{pmatrix}
      0 & \beta
      \\
      -1 & \alpha
   \end{pmatrix}^n
   _\bullet
   g_0^\mathrm{L}(z),
   \label{eq:An}
\end{align}
with
\begin{align}
   x
   =&
   \begin{pmatrix}
      a & b
      \\
      c & d
   \end{pmatrix}
   _\bullet
   y
   \equiv
   (ay+b)/(cy+d)
   ,
   \\
   g_0^\mathrm{L}(z)
   =&
   -t^2, 
   \\
   \begin{pmatrix}
      0 & \beta
      \\
      -1 & \alpha
   \end{pmatrix}^n
   =&
   \frac{2^{-n-1}}{\gamma}
   \begin{pmatrix}
      -4\beta y_{n-1} & 2\beta y_n
      \\
      -2y_n & y_{n+1}
   \end{pmatrix},
   \\
   y_n
   =&
   (\alpha+\gamma)^n
   -
   (\alpha-\gamma)^n,
   \\
   \alpha
   =&
   z^2-\mu^2/4-t^2,
   \\
   \beta
   =&
   z^2t^2/4,
   \\
   \gamma
   =&
   \sqrt{\alpha^2-4\beta},
\end{align}
where $g_0^\mathrm{L}(z)$ corresponds to the vacuum state [$f_0^\mathrm{L}(z)=0$ in Eq.~(\ref{eq:f_n})].

The surface Green's function at the rightmost site $G_1^{\mathrm{L},\infty}(z)$ is given 
by taking the limit of $n\rightarrow\infty$ as
\begin{align}
   G_1^{\mathrm{L},\infty}(z)
   \equiv&
   \lim_{n\rightarrow\infty}
   [G_{n}(z)]_{1,1}
   \nonumber\\
   =&
   \frac{1}{z^2-\frac{\mu^2}{4}-2zf^\mathrm{L}_\infty(z)}
   \begin{pmatrix}
      z-\frac{\mu}{2}-f^\mathrm{L}_\infty & -f^\mathrm{L}_\infty
      \\
      -f^\mathrm{L}_\infty & z+\frac{\mu}{2}-f^\mathrm{L}_\infty
   \end{pmatrix},
   \label{eq:surface_G_kitaev}
\end{align}
with
\begin{align}
   f^\mathrm{L}_\infty(z) \equiv&
   \lim_{n\rightarrow\infty}f^\mathrm{L}_n(z)
   =
   \lim_{n\rightarrow\infty}\frac{g^\mathrm{L}_n(z)+t^2}{2z},
   \\
   g^\mathrm{L}_\infty(z)
   \equiv&
   \lim_{n\rightarrow\infty}
   g^\mathrm{L}_n(z)
   \nonumber\\
   =&
   \begin{cases}
      \frac{1}{2}
      (\alpha-\gamma)
      \:\:\mathrm{for}[|(\alpha-\gamma)/(\alpha+\gamma)|<1],
      \\
      \frac{1}{2}
      (\alpha+\gamma)
      \:\:\mathrm{for}[|(\alpha-\gamma)/(\alpha+\gamma)|>1].
   \end{cases}
\end{align}
\subsection{Local Green's function for arbitrary site for $t=\Delta$\label{sec:sG_arbit_kitaev}}

The recurrence relation for the Green's function at leftmost site is given by
\begin{align}
   [G_{n+1}(z)]_{n+1,n+1}
   =&
   \frac{1}{z-\hat{u}-\hat{t} [G_{n}(z)]_{n,n}\hat{t}^\dag},
   \label{eq:rec_kitaev_left}
\end{align}
and we get the local Green's function at the leftmost site:
\begin{align}
   &[G_{n+1}(z)]_{n+1,n+1}
   \nonumber\\
   =&
   \frac{1}{z^2-\frac{\mu^2}{4}-2zf^\mathrm{R}_{n}(z)}
   \begin{pmatrix}
      z-\frac{\mu}{2}-f^\mathrm{R}_{n}(z) & f^\mathrm{R}_{n}(z)
      \\
      f^\mathrm{R}_{n}(z) & z+\frac{\mu}{2}-f^\mathrm{R}_{n}(z)
   \end{pmatrix},
   \label{eq:green_R}
\end{align}
where 
\begin{align}
   f_{n}^\mathrm{R}(z)
   =
   \sum_{l,m=1,2}(-1)^{l+m}\{[G_{n}(z)]_{n,n}\}_{lm}.
   \label{eq:fR}
\end{align}
$g_{n}^\mathrm{R}(z)\equiv2zf_{n}^\mathrm{R}(z)-t^2$ satisfies the same equation 
as Eq.~(\ref{eq:An}).
It is noted that the difference between Eq.~(\ref{eq:rec_kitaev}) and Eq.~(\ref{eq:green_R})
is the sign of the offdiagonal element.
Let a $2\times2$ matrix $[G_{j}^\infty(z)]$ be
\begin{align}
   G_{j}^\infty(z)
   \equiv&
   \lim_{n\rightarrow\infty}[G_{n}(z)]_{j,j}.
\end{align}
$G_1^\infty(z)$ is given by Eq.~(\ref{eq:surface_G_kitaev}) [$G_1^\infty(z)=G_1^{\mathrm{L},\infty}(z)$].
The local Green's function for an arbitrary site other than the surface ($j>1$) is
obtained by using following equations.
\begin{align}
   {[G_{n}(z)]}_{j,j}
   =&
   \frac{1}{%
      {[G_{j}(z)]}_{j,j}^{-1}
      -
      \hat{t}^\dag
      {[G_{n-j}]}_{1,1}
      \hat{t}
   },
   \\
   {[G_{n}(z)]}_{j+1,j}
   =&
   [G_{n-j}(z)]_{1,1}
   \hat{t}
   [G_n(z)]_{j,j},
   \\
   {[G_{n}(z)]}_{j,j+1}
   =&
   [G_n(z)]_{j,j}
   \hat{t}^\dag
   [G_{n-j}(z)]_{1,1}.
\end{align}
Then the local Green's function for $t=\Delta$ is
\begin{align}
   [G_{j}^\infty(z)]_{11}
   =&
   \alpha'
   \left[
      \frac{\delta_-(1-\delta_{+} X_{j-1})}
      {1-X_{j-1}(\delta_++\delta_-)}
      +
      f_\infty^\mathrm{L}(z)
   \right],
   \\
   [G_{j}^\infty(z)]_{22}
   =&
   \alpha'
   \left[
      \frac{\delta_+(1-\delta_{-} X_{j-1})}
      {1-X_{j-1}(\delta_++\delta_-)}
      +
      f_\infty^\mathrm{L}(z)
   \right],
   \\
   [G_{j}^\infty(z)]_{12}
   =&
   [G_{j}^\infty(z)]_{21}
   \nonumber\\
   =&
   \alpha'
   \left[
      \frac{\delta_+\delta_- X_{j-1}}
      {1-X_{j-1}(\delta_++\delta_-)}
      -
      f_\infty^\mathrm{L}(z)
   \right],
   \label{eq:Green_func_G12_kitaev_Dt}
\end{align}
with
\begin{align}
   \alpha'
   =&
   \frac{1}{\alpha-g_\infty^\mathrm{L}(z)},
   \\
   \delta_\pm
   =&
   z\pm\mu/2-2f_\infty^\mathrm{L}(z),
   \\
   X_{j}
   =&
   \frac{t^2}{2}
   \frac{z-2f_{j}^\mathrm{L}}{(\alpha-g_{j}^\mathrm{L})(\alpha-g_{\infty}^\mathrm{L})}.
\end{align}
From the above definition, $G_{j}^\infty(z)$ depends on the site index $j$ through $X_j$.

\subsection{SBBC with $t=\Delta$ and $\mu=0$\label{sec:sG_mu0_kitaev}}
For $\mu=0$, the local Green's function is easily calculated and its value does not depend
on the site other than the surface.

From Eqs.~(\ref{eq:f_n}) and (\ref{eq:rec_kitaev2}) with $\mu=0$,
$f_n^\mathrm{L}(z)$ is
\begin{align}
   f_n^\mathrm{L}(z)
   =&
   \frac{t^2}{2z},
\end{align}
and it does not depend on $n$.
Then the local Green's function at the rightmost site is calculated from Eq.~(\ref{eq:rec_kitaev}) as
\begin{align}
   G_1^{\mathrm{L},{\infty}}(z)
   =&
   \frac{1}{2z}
   \frac{1}{z^2-t^2}
   \begin{pmatrix}
      2z^2-t^2 & -t^2
      \\
      -t^2 & 2z^2-t^2
   \end{pmatrix}.
\end{align}
The local Green's function for the arbitrary site other than the surface ($j>1$)
is calculated by using the recursive Green's function method and
they have the same value given by
\begin{align}
   G_{j}^\infty(z)
   =&
   \frac{z}{z^2-t^2}
   \begin{pmatrix}
      1 &0
      \\
      0& 1
   \end{pmatrix}.
\end{align}

Then we show the SBBC for the Kitaev chain with $t=\Delta$ and $\mu=0$.
The chiral operator for the Kitaev chain with the semi-infinite system is
\begin{align}
   \Gamma 
   =&
   \begin{pmatrix}
      \ddots
      \\
      & \tau_1
      \\
      && \tau_1
   \end{pmatrix}
\end{align}
with the Pauli matrix $\tau_1$.
$F_\mathrm{edge}^\mathrm{odd}(z)$
can be exactly evaluated as
\begin{align}
   F_\mathrm{edge}^\mathrm{odd}(z)
   =&
   \mathrm{Tr}\Gamma G(z)
   \nonumber\\
   =&
   \mathrm{Tr}
   \begin{pmatrix}
      \ddots
      \\
      & \tau_1
      \\
      && \tau_1
   \end{pmatrix}
   \begin{pmatrix}
      \ddots
      \\
      & G_{2}^\infty(z)
      \\
      && G_{1}^\infty(z)
   \end{pmatrix}
   \nonumber\\
   =&
   \mathrm{Tr}
   \begin{pmatrix}
      \ddots
      \\
      &0
      \\
      &&0
      \\
      &&& \frac{1}{2z}\frac{-t^2}{z^2-t^2}
      \\
      &&&& \frac{1}{2z}\frac{-t^2}{z^2-t^2}
   \end{pmatrix}
   \nonumber\\
   =&
   \frac{1}{z}
   \frac{-t^2}{z^2-t^2}.
   \label{eq:BEC_kitaev_L}
\end{align}

$w_\mathrm{bulk}(z)$ is given by Eq.~(\ref{eq:w_bulk_Kitaev_gen}) as
\begin{align}
   w_\mathrm{bulk}(z)
   =&
   \frac{-t^2}{z^2-t^2},
   \label{eq:BEC_kitaev_R}
\end{align}
where we set $t=\Delta$ and $\mu=0$.
Then the SBBC is realized as
\begin{align}
   F_\mathrm{edge}^\mathrm{odd}(z)
   =
   w_\mathrm{bulk}(z)/z.
\end{align}
\subsection{One-dimensional $p$-wave superconductor in the continuum model\label{sec:kitaev_cont}}
\subsubsection{Green's function for semi-infinite system}
In the main text, we assume that the edge is located on the right side, however,
if the edge is located on the left side, the sign of $F_\mathrm{edge}^\mathrm{odd}(z)$ becomes opposite\cite{sato-tanaka10}.
In this subsection, we assume that the edge is located on the left side.

First, we show the Green's function of 
a semi-infinite spin-singlet $s$-wave superconductor. 
If we focus on one spin-sector, the Hamiltonian is given by 
\begin{align}
   H=(\xi_{k}\tau_{3} + \Delta \tau_{1})\Theta(x).
\end{align}
The resulting Green's function is~\cite{McMillan,TK96a,Lu20150246}
\begin{align}
   &
   \left(
      \frac{m}{\hbar^{2} \mathrm{i}}
   \right) 
   \frac{1}{1 - \bar{\Gamma}^{2}}
   \left\{
      \frac{1}{k^{+}}
      \begin{pmatrix}
         1 & \bar{\Gamma} \\
         \bar{\Gamma} & \bar{\Gamma}^{2}
      \end{pmatrix}
      \left[%
         e^{\mathrm{i}k^{+} | x -x' |}
         - 
         e^{\mathrm{i}k^{+} ( x +x' )}
      \right]
      \right.
      \nonumber\\
      & \hspace{0.5cm}
      +
      \left.
      \frac{1}{k^{-}}
      \begin{pmatrix}
         \bar{\Gamma}^{2} & \bar{\Gamma} \\
         \bar{\Gamma} & 1
      \end{pmatrix}
      \left[
         e^{-\mathrm{i}k^{-} | x -x' |}
         - 
         e^{-\mathrm{i}k^{-} ( x +x' )}
      \right]
   \right\} 
\end{align}
where $k^{+}$ and $k^{-}$ denote 
\begin{align}
   k^{\pm}
   =&
   \sqrt{
      \frac{2m}{\hbar^{2}}
      \left(
         E_\mathrm{F} \pm \Omega
      \right)
   }, 
   \\
   \Omega
   =&
   \sqrt{E^{2} -\Delta^{2}}. 
\end{align}
Here, $\Omega$ expresses 
\begin{align}
   \Omega 
   \equiv&
   \lim_{\delta \rightarrow 0} 
   \sqrt{
      \left(
         \omega + \mathrm{i}\delta
      \right)^{2} 
      - 
      \Delta_{0}^{2}
   }
   \nonumber\\
   =& 
   \left\{
      \begin{array}{ll}
         \sqrt{\omega^{2} - \Delta_{0}^{2}} & \omega  \geq \Delta_{0} \\
         \mathrm{i} \sqrt{\Delta_{0}^{2} - \omega^{2}} &  -\Delta_{0} \leq \omega \leq \Delta_{0} \\ 
         -\sqrt{\omega^{2} - \Delta_{0}^{2}} & \omega  \leq -\Delta_{0} 
      \end{array}
   \right.
   \label{Omega}
\end{align}
with infinitesimal small positive number $\delta$. 
In the above, $\bar{\Gamma}$ is given by 
\begin{align}
   \bar{\Gamma}=\frac{\Delta}{E + \Omega}. 
\end{align}

In usual case, the magnitude of $E_\mathrm{F}$ is much larger than 
$\Omega$. 
Then, $k_{\pm}$ can be approximated as  
\begin{align}
   k^{\pm} \sim& k_\mathrm{F} + \gamma, 
   \\
   \gamma=& \frac{\Omega}{\hbar v_\mathrm{F}}, 
\end{align}
Then, 
both  the 12 and 21 components of this $2 \times 2$ matrix is given by 
\begin{align}
   &
   \left(
      \frac{m}{\mathrm{i}k_\mathrm{F}\hbar^{2}}
   \right)
   \frac{\Delta}{\Omega}
   \left\{
      e^{\mathrm{i}\gamma | x-x' |} 
      \cos[k_\mathrm{F}(x-x')]
   \right.
   \nonumber\\
   &\hspace{2.2cm}
   \left.
      - 
      e^{\mathrm{i}\gamma (x + x')}
      \cos[k_\mathrm{F} (x + x')]
   \right\}
\end{align}

In Matsubara representation, it is written as 
\begin{align}
   &
   \left(
      \frac{-1}{\hbar v_\mathrm{F}}
   \right)
   \frac{\Delta}{\Omega_{n}}
   \left\{
      e^{-\gamma_{n} | x -x' | }
      \cos[k_\mathrm{F}(x -x')]
   \right.
   \nonumber\\
   &\hspace{2.2cm}
   \left.
      -
      e^{-\gamma_{n}(x+x') }
      \cos[k_\mathrm{F} (x +x')]
   \right\}
\end{align}
with $\gamma_{n}=\sqrt{\omega_{n}^{2} + \Delta^{2}}/(\hbar v_\mathrm{F})$. 
We can easily understand that only the even-frequency pairing exists. 
There is no induced odd-frequency pairings. 
In the spin-singlet $s$-wave superconductor semi-infinite system, 
without considering spatial dependence of pair potential, 
there is no induced odd-frequency pairing. \par

Next, we focus on the Green's function of the Kitaev chain 
within quasi-classical approximations. 
The pair potential is give by~\cite{sato-tanaka10} 
\begin{align}
   \Delta(k)=\Delta \frac{k}{\sqrt{k^{2}}}. 
\end{align}
In the bulk, the Hamiltonian is given by 
\begin{align}
   H=\xi_{k}\tau_{3} + \Delta(k)\tau_{1}.
\end{align}
The Green's function in the bulk is given by 
\begin{align}
   &
   \left(
      \frac{m}{ik_\mathrm{F}\hbar^{2}}
   \right)
   \left(
      1-\bar{\Gamma}^{2}
   \right)^{-1}
   e^{\mathrm{i}\gamma | x -x' |}
   \nonumber\\
   &
   \times 
   \left[
      e^{\mathrm{i}k_\mathrm{F} | x -x' |}
      \begin{pmatrix}
         1 & 0 \\
         0 & \bar{\Gamma}^{2}
      \end{pmatrix}
      + 
      e^{-\mathrm{i}k_\mathrm{F} | x -x' |}
      \begin{pmatrix}
         \bar{\Gamma}^{2} & 0 \\
         0 & 1
      \end{pmatrix}
   \right.
   \nonumber\\
   &\hspace{1cm}
   \left.
      + 
      2 \mathrm{i} \bar{\Gamma} \sin[k_\mathrm{F}(x-x')]
      \begin{pmatrix}
         0 & 1 \\
         1 & 0
      \end{pmatrix}
   \right].
\end{align}

Let us consider a semi-infinite system. 
The results are given as follows. 
\begin{widetext}
\begin{align}
   &
   \left(
      \frac{m}{\mathrm{i}k_\mathrm{F}\hbar^{2}}
   \right)
   \left(
      1-\bar{\Gamma}^{2}
   \right)^{-1}
   \left\{
      e^{\mathrm{i}\gamma | x -x' |}
      \left[
         e^{\mathrm{i}k_\mathrm{F} | x -x' |}
         \begin{pmatrix}
            1 & 0 
            \\
            0 & \bar{\Gamma}^{2}
         \end{pmatrix}
         + 
         e^{-\mathrm{i}k_\mathrm{F} | x -x' |}
         \begin{pmatrix}
            \bar{\Gamma}^{2} & 0 
            \\
            0 & 1
         \end{pmatrix}
         + 
         2 \mathrm{i} \bar{\Gamma} \sin[k_\mathrm{F}(x-x')]
         \begin{pmatrix}
            0 & 1 
            \\
            1 & 0
         \end{pmatrix}
      \right]
   \right.
   \nonumber\\
   &
   \left.
      + 
      e^{\mathrm{i}\gamma(x+x')}
      \frac{1-\bar{\Gamma}^{2}}{1 + \bar{\Gamma}^{2}}
      \begin{pmatrix}
         -
         e^{\mathrm{i}k_\mathrm{F}(x+x')}
         + 
         \bar{\Gamma}^{2}
         e^{-\mathrm{i}k_\mathrm{F}(x+x')}
         & 
         2\bar{\Gamma} \cos[k_\mathrm{F}(x+x')] 
         \\
         -
         2\bar{\Gamma} \cos[k_\mathrm{F}(x+x')] 
         &
         \bar{\Gamma}^{2} 
         e^{\mathrm{i}k_\mathrm{F}(x+x')}
         - 
         e^{-\mathrm{i}k_\mathrm{F}(x+x')}
      \end{pmatrix}
      - 
      e^{i\gamma(x+x')}
      \frac{2\bar{\Gamma}}{1 + \bar{\Gamma}^{2}}
   \right.
   \nonumber\\
   &
   \left.
      \times
      \begin{pmatrix}
         2\bar{\Gamma} \cos[k_\mathrm{F}(x-x')]
         & 
         (1 - \bar{\Gamma}^{2})\cos[k_\mathrm{F}(x-x')]
         + \mathrm{i}(1 + \bar{\Gamma}^{2})\sin[k_\mathrm{F}(x-x')]
         \\
         -(1 - \bar{\Gamma}^{2})\cos[k_\mathrm{F}(x-x')]
         + \mathrm{i}(1 + \bar{\Gamma}^{2})\sin[k_\mathrm{F}(x-x')]
         &
         2\bar{\Gamma} \cos[k_\mathrm{F}(x-x')] 
      \end{pmatrix}
   \right\}.
\end{align}
We further focus on the 12 and 21 components of this matrix. 
The 12 component of this matrix can be expressed as  
\begin{align}
   \left(
      \frac{m}{ik_\mathrm{F}\hbar^{2}}
   \right)
   \left\{
      \frac{\Delta}{\Omega}\mathrm{i} \sin[k_\mathrm{F}(x-x')]
      \left[
         e^{\mathrm{i}\gamma \mid x -x' \mid }
         - 
         e^{\mathrm{i}\gamma (x+x')}
      \right] 
      -
      2\frac{\Delta}{E}
      e^{\mathrm{i}\gamma(x+x')}
      \sin(k_\mathrm{F}x)\sin(k_\mathrm{F}x')
   \right\}.
\end{align}
On the other hand, the 21 component is given by 
\begin{align}
   \left(
      \frac{m}{ik_\mathrm{F}\hbar^{2}}
   \right)
   \left\{
      \frac{\Delta}{\Omega}\mathrm{i} \sin[k_\mathrm{F}(x-x')]
      \left[
         e^{\mathrm{i}\gamma | x -x' |}
         - 
         e^{\mathrm{i}\gamma (x+x')}
      \right] 
      +
      2\frac{\Delta}{E}
      e^{\mathrm{i}\gamma(x+x')}
      \sin(k_\mathrm{F}x)\sin(k_\mathrm{F}x')
   \right\}.
\end{align}
\end{widetext}
\subsubsection{SBBC}
   Let $\tilde{F}_\mathrm{edge}^\mathrm{odd}(z)$ be
   \begin{align}
      \tilde{F}_\mathrm{edge}^\mathrm{odd} (z)
      =&
      \sum_j 
      \langle j| \Gamma G(z) |j\rangle,
   \end{align}
   where the edge is located on the left side.
   Since the edge is located on the right side 
   in the definition of $F_\mathrm{edge}^\mathrm{odd}(z)$,
   the sign of $\tilde{F}_\mathrm{edge}^\mathrm{odd}(z)$
   and $F_\mathrm{edge}^\mathrm{odd}(z)$ are opposite:
   \begin{align}
      \tilde{F}_\mathrm{edge}^\mathrm{odd}(z)
      =&
      -F_\mathrm{edge}^\mathrm{odd}(z).
   \end{align}
The chiral operator is
\begin{align}
   \Gamma = \tau_2,
\end{align}
and $\tilde{F}_\mathrm{edge}^\mathrm{odd}(z)$ is
\begin{align}
   \tilde{F}_\mathrm{edge}^\mathrm{odd}(E)
   =&
   \frac{m}{ik_\mathrm{F}\hbar^2}
   \int_0^\infty dx
   (-4\mathrm{i})
   \frac{\Delta}{E}
   e^{2\mathrm{i}\gamma x}\sin^2(k_\mathrm{F}x)
   \nonumber\\
   =&
   -
   \frac{m}{\mathrm{i}k_\mathrm{F}\hbar^2}
   \frac{i}{2}
   \frac{\Delta}{E}
   \left[
      \frac{1}{\mathrm{i}(\gamma-k_\mathrm{F})}
      +
      \frac{1}{\mathrm{i}(\gamma-k_\mathrm{F})}
      -
      \frac{2}{\mathrm{i}\gamma}
   \right].
\end{align}
We suppose $k_\mathrm{F}\gg\gamma$.
Then we obtain
\begin{align}
   \tilde{F}_\mathrm{edge}^\mathrm{odd}(z)
   =&
   \frac{m}{ik_\mathrm{F}\hbar^2}
   \frac{\mathrm{i}}{2}
   \frac{\Delta}{z}
   \frac{2}{\mathrm{i}\gamma}
   \nonumber\\
   =
   &
   -
   \frac{\Delta}{z\sqrt{\Delta^2-z^2}}.
   \label{eq:continuum_F}
\end{align}

The Hamiltonian for the bulk is given by
\begin{align}
   {\cal H}_\mathrm{bulk}
   =&
   \sum_k
   C_k^\dag
   \begin{pmatrix}
      \xi_k & \Delta(k)
      \\
      \Delta(k) & -\xi_k
   \end{pmatrix}
   C_k,
\end{align}
where $\xi_k=\hbar^2k^2/2m-\mu$ and $\Delta(k)=\Delta k/\sqrt{k^2}$ for $k\neq0$,
$\Delta(k=0)=0$ and $C_k=(c_k,c_{-k}^\dag)^\mathrm{T}$.
The Green's function is
\begin{align}
   g(z,k)
   =&
   \frac{1}{z^2-\xi_k^2-\Delta^2(k)}
   \begin{pmatrix}
      z+\xi_k & \Delta(k)
      \\
      \Delta(k) & z-\xi_k
   \end{pmatrix}.
\end{align}
Then we get
\begin{align}
   w_\mathrm{bulk}(z)
   =&
   \frac{\mathrm{i}}{4\pi}
   \int_{-\infty}^\infty \mathrm{d}k\:
   \mathrm{tr}\Gamma g(z,k)\partial_k g^{-1}(z,k)
   \nonumber\\
   =&
   \frac{\Delta}{\sqrt{\Delta^2-z^2}}.
   \label{eq:continuum_w}
\end{align}

   Finally, we get
   \begin{align}
      -\tilde{F}_\mathrm{edge}^\mathrm{odd}(z)
      =&
      F_\mathrm{edge}^\mathrm{odd}(z)
      =
      w_\mathrm{bulk}(z)/z.
   \end{align}
\subsection{Numerical results of SBBC for Kitaev chain\label{sec:SBEC_mun0_kitaev}}
\subsubsection{$\Delta/t=1$}
\begin{figure*}[t]
   \centering
   \includegraphics[width = 17cm]{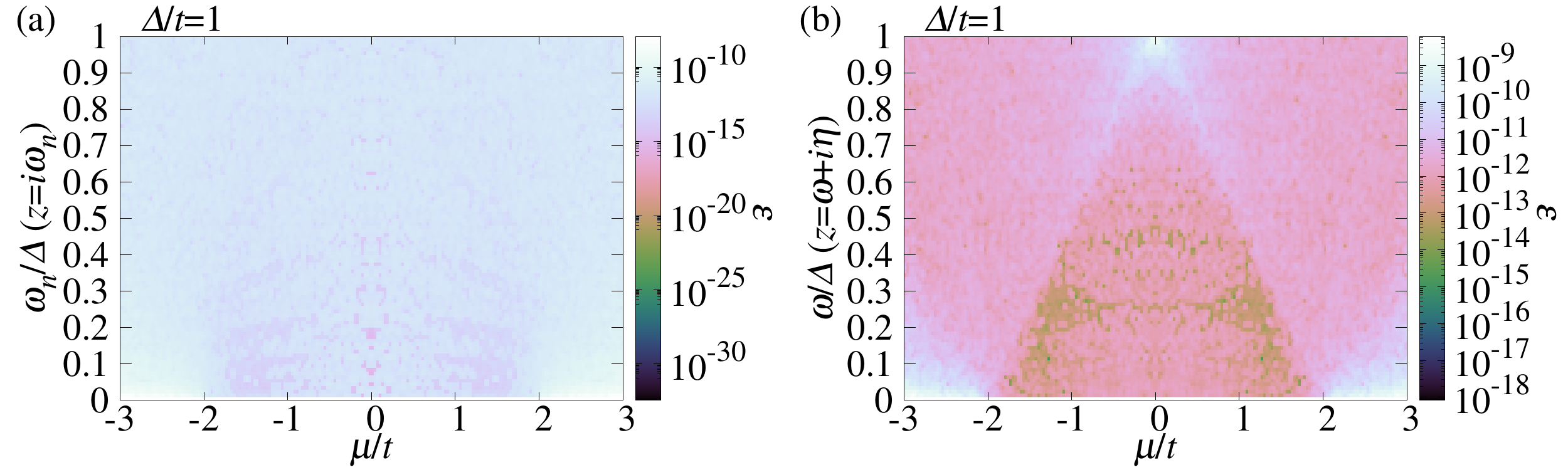}
   \caption{%
      $\varepsilon$ for the Kitaev chain with $\Delta/t=1$ is plotted as functions of 
      (a) $\mu/t$ and $\omega_n/\Delta$ ($z=\mathrm{i}\omega_n$) and 
      (b) $\mu/t$ and $\omega/\Delta$ ($z=\omega+\mathrm{i}\eta$ with $\eta=10^{-2}$).
   }%
   \label{fig:kitaev_D1_diff_theta1_2}
\end{figure*}

We numerically calculate $F_\mathrm{edge}^\mathrm{odd}(z)$ by using 
Eq.~(\ref{eq:Green_func_G12_kitaev_Dt}) and compare it with 
$w_\mathrm{bulk}(z)/z$ [Eq.~(\ref{eq:kitaev_w_exact})]
for $t=\Delta$ and $\mu\neq0$.
When we calculate $\mathrm{Tr}_j$ in $F_\mathrm{edge}^\mathrm{odd}(z)$, we take $10^4$ sites from the surface.
In Fig.~\ref{fig:kitaev_D1_diff_theta1_2}, we show 
\begin{align}
   \varepsilon
   =&
   \frac{|F_\mathrm{edge}^\mathrm{odd}(z)-w_\mathrm{bulk}(z)/z|}{|w_\mathrm{bulk}(z)/z|}
\end{align}
as functions of $\mu/t$ and $z/\Delta$ for $\Delta/t=1$.
In Fig.~\ref{fig:kitaev_D1_diff_theta1_2}(a), we set $z=i\omega_n$ and 
in (b), we set $z=\omega+i\eta$ ($\eta/t=10^{-2}$).
The order of $\varepsilon$ is less than $10^{-8}$ for the both cases.

\subsubsection{$\Delta/t=0.05$\label{sec:SBEC_arb_D_kitaev}}
\begin{figure*}[t]
   \centering
   \includegraphics[width = 17cm]{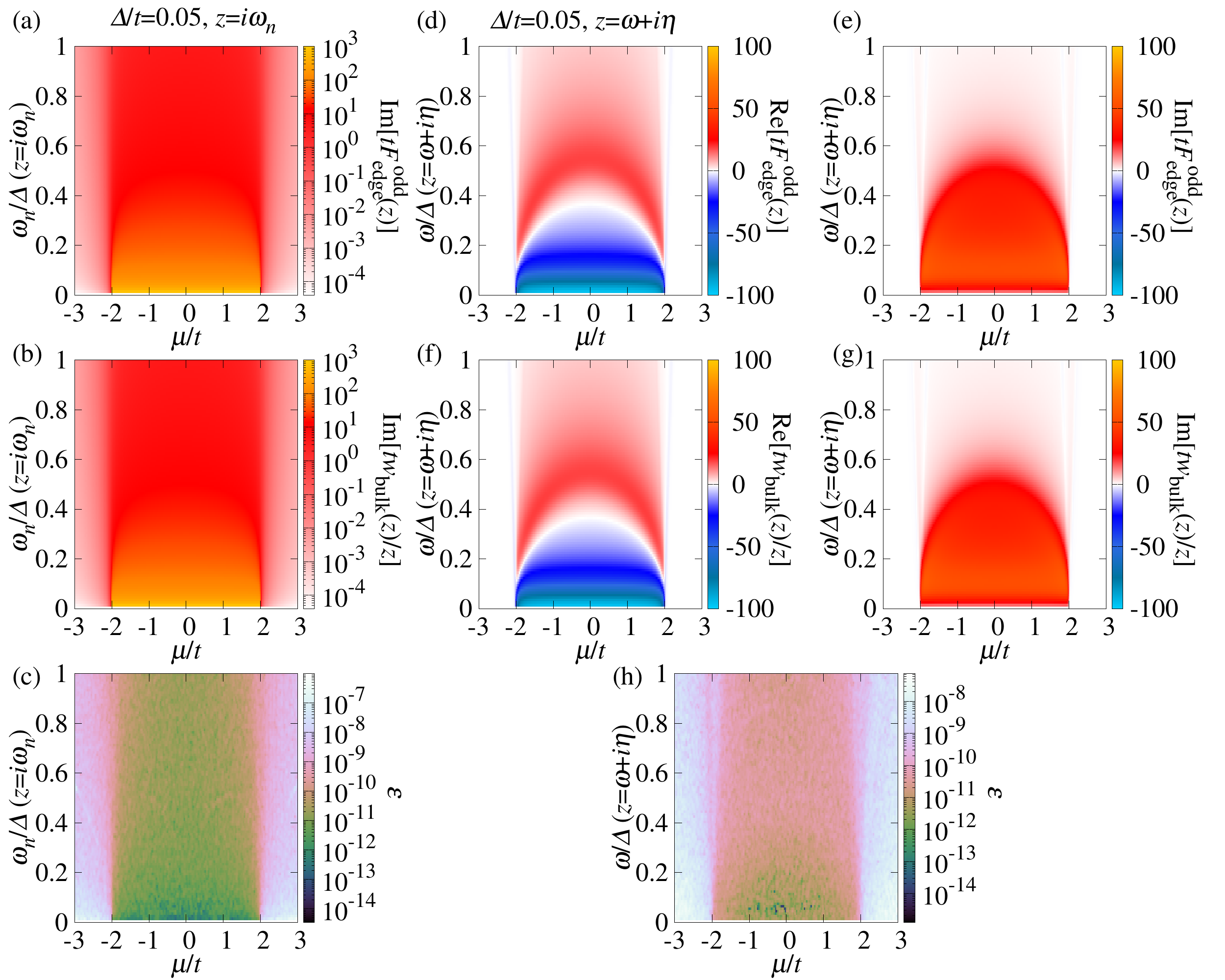}
   \caption{%
      (a) Im$[F_\mathrm{edge}^\mathrm{odd}(z)]$,
      (b) Re$[w_\mathrm{bulk}(z)/z]$ and
      (c) $\varepsilon$ for Kitaev chain with $\Delta/t=0.05$ are plotted as functions of 
      $\mu/t$ and $\omega_n/\Delta$ ($z=\mathrm{i}\omega_n$).
      (d) Re$[F_\mathrm{edge}^\mathrm{odd}(z)]$,
      (e) Im$[F_\mathrm{edge}^\mathrm{odd}(z)]$,
      (f) Re$[w_\mathrm{bulk}(z)/z]$,
      (g) Im$[w_\mathrm{bulk}(z)/z]$ and
      (h) $\varepsilon$ for Kitaev chain with $\Delta/t=0.1$ are plotted as functions of
      $\mu/t$ and $\omega/\Delta$ ($z=\omega+\mathrm{i}\eta$ with $\eta=10^{-2}$).
   }%
   \label{fig:kitaev_D0_1}
\end{figure*}

We numerically calculate $F_\mathrm{edge}^\mathrm{odd}(z)$ by using the recursive Green's function method~\cite{Umerski} 
and $w_\mathrm{bulk}(z)/z$ by using Eq.~(\ref{eq:w_bulk_Kitaev_gen}) with $\Delta/t=0.05$.
When we calculate $\mathrm{Tr}_j$ in $F_\mathrm{edge}^\mathrm{odd}(z)$, we take $10^5$ sites from the surface and
when we calculate integral in $w_\mathrm{bulk}(z)$, we also take $10^5$ wave numbers.
In Fig.~\ref{fig:kitaev_D0_1}, we show 
(a) $F_\mathrm{edge}^\mathrm{odd}(z)$, 
(b) $w_\mathrm{bulk}(z)/z$ and
(c) $\varepsilon$ as functions of $\mu/t$ and $\omega_n/\Delta$ ($z=i\omega_n$) for $\Delta/t=0.05$.
We also show 
(d) Re$[F_\mathrm{edge}^\mathrm{odd}(z)]$, 
(e) Im$[F_\mathrm{edge}^\mathrm{odd}(z)]$, 
(f) Re$[w_\mathrm{bulk}(z)/z]$,
(g) Im$[w_\mathrm{bulk}(z)/z]$ and
(h) $\varepsilon$ as functions of $\mu/t$ and $\omega/\Delta$ ($z=\omega+i\eta$) for $\Delta/t=0.05$.
The order of $\varepsilon$ is less than $10^{-6}$ for $z=i\omega_n$ and 
$10^{-7}$ for $z=\omega+i\eta$.
As can be seen in Fig.~\ref{fig:kitaev_D0_1}, $\varepsilon$ becomes large
where $F_\mathrm{edge}^\mathrm{odd}(z)$ and $w_\mathrm{bulk}(z)/z$ are small.

\subsection{Spatial modulation of gap function and chemical potential\label{sec:SBEC_kitaev_sp_mod}}
\begin{figure*}[t]
   \centering
   \includegraphics[width = 17cm]{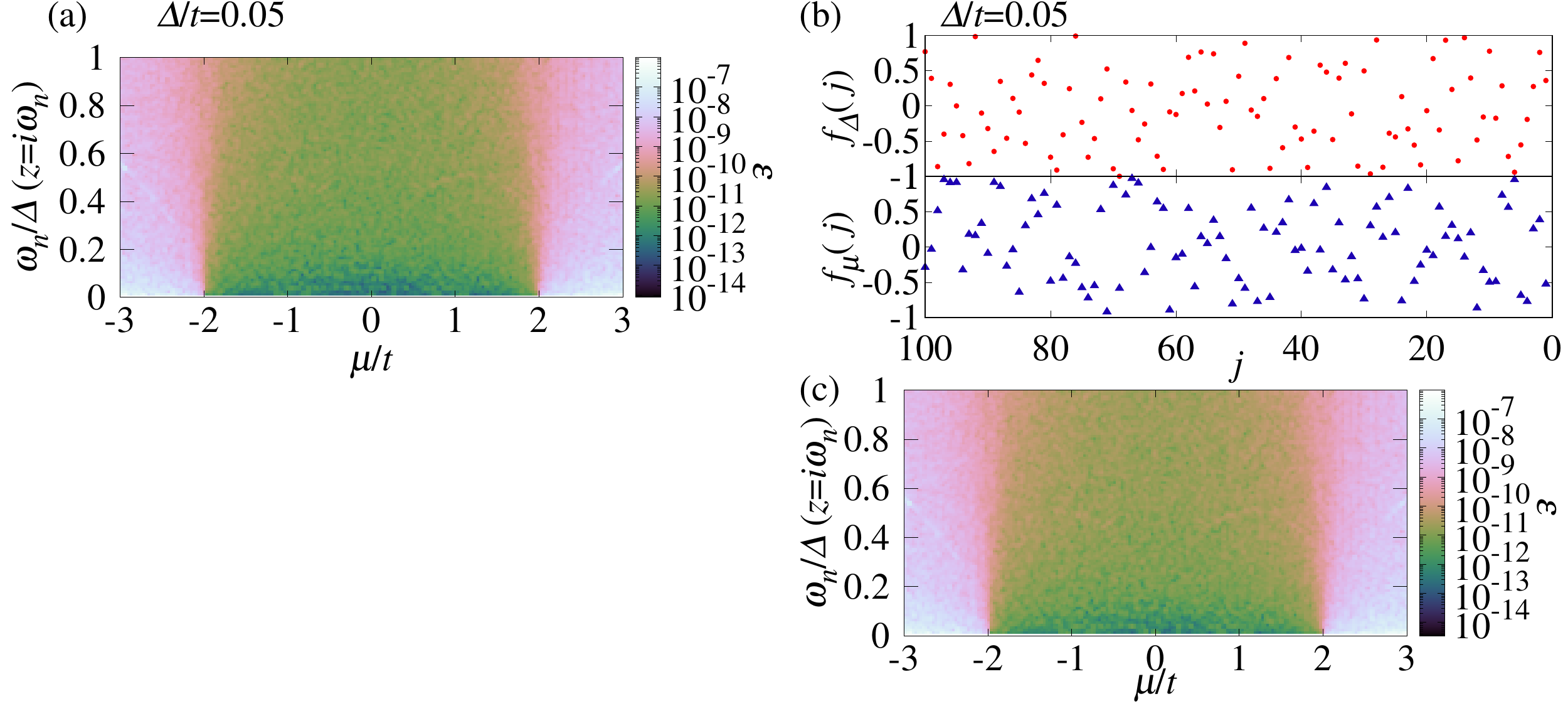}
   \caption{%
      (a) $\varepsilon$ with surface modulation of the gap function [$\tanh(j\Delta/t)$]
      is plotted as functions of $\mu/t$ and $\omega_n/\Delta$ for $\Delta/t=0.05$.
      (b) randomly chosen $f_\Delta(j)$ and $f_\mu(j)$ are plotted as a function of $j$.
      (c) $\varepsilon$ with surface modulation
      shown in (b)
      is plotted as functions of $\mu/t$ and $\omega_n/\Delta$ for $\Delta/t=0.05$.
   }%
   \label{fig:kitaev_D0_1_tanh}
\end{figure*}

In this subsection we show that the SBBC holds even if we introduce spatial modulation of
the gap function and the chemical potential near the surface.
In Fig.~\ref{fig:kitaev_D0_1_tanh}(a), we show $\varepsilon$ where gap function is the form $\tanh(j\Delta/t)$.
In Fig.~\ref{fig:kitaev_D0_1_tanh}(c), we show $\varepsilon$ where gap function and chemical potential
are randomly chosen.
In these cases the Hamiltonian with semi-infinite system is
\begin{align}
   {\cal H}
   =&
   -t\sum_j
   \left(
      c_j^\dag c_{j+1}
      +\mathrm{H.c.}
   \right)
   +
   \Delta\sum_j
   f_\Delta(j)
   \left(
      c_{j+1}^\dag c_j^\dag
      +\mathrm{H.c.}
   \right)
   \nonumber\\
   &
   -
   \sum_j
   \left[
      \mu + tf_\mu(j)
   \right]
   n_j.
   \label{eq:kitaev_Hamiltonian_surface_mod}
\end{align}
   Note that this Hamiltonian has chiral symmetry.
In Fig.~\ref{fig:kitaev_D0_1_tanh}(a), we set $f_\Delta(j)=\tanh(j\Delta/t)$ and $f_\mu(j)=1$
and in Fig.~\ref{fig:kitaev_D0_1_tanh}(c), $|f_\Delta(j)|<1$ and $|f_\mu(j)|<1$ are random values
[These values are shown in Fig.~\ref{fig:kitaev_D0_1_tanh}(b)]
for $j\leq100$ and $f_\Delta(j)=1$ and $f_\mu(j)=0$ for $j>100$.

We can conclude from the numerical results in figures that the SBBC holds even 
when the spatial modulation near the surface is taken into account.

\subsection{Exact formula of $\chi$ for Kitaev chain
   with any $\Delta/t>0$
\label{sec:chi_exact_kitaev}}
To obtain $\chi$ away from the transition point,
we analytically calculate $\chi$ for $\mu>0$.
$\chi$ for $\mu<0$ can be obtained in a similar manner.
$w_\mathrm{bulk}(z)$ for arbitrary $\Delta/t$ ($\Delta/t>0$) is
given by Eq.~(\ref{eq:w_bulk_Kitaev_gen}).
Let $x=e^{\mathrm{i}k}$ be a complex number then
$w_\mathrm{bulk}(z)$ can be written as
\begin{widetext}
   \begin{align}
      w_\mathrm{bulk}(z)
      =&
      \frac{1}{2\pi}
      \int_{C}
      \frac{\mathrm{d}x}{\mathrm{i}x}
      \frac{%
         -t\Delta
         -
         \frac{\mu\Delta}{2}(x+x^{-1})/2
      }
      {z^2+\Delta^2(x-x^{-1})^2/4-[-t(x+x^{-1})/2-\mu/2]^2},
\label{eq:w_kitaev}
   \end{align}
\end{widetext}
where the integral path $C$ is a unit circle in the complex
plane.
Then $\chi$ is obtained by the second derivative 
of $w_\mathrm{bulk}(z)$ as
\begin{align}
   \chi
   =&
   -
   \frac{1}{2\pi \mathrm{i}}
   \int_{C}
   \mathrm{d}x
   \frac{%
      \left[
         -t\Delta x
         -
         \frac{\mu\Delta}{2}(x^2+1)/2
      \right]x^2
   }%
   {%
      \left\{
         \frac{\Delta^2(x^2-1)^2}{4}
         -
         \left[
            -t(x^2+1)/2-\mu x/2
         \right]^2
      \right\}^2
   }.
\end{align}
There are three cases.
\\
i) $\mu=0$ and $\Delta/t=1$

There is a pole at $x=0$ and $\chi=1$ (See Eq.~\ref{eq:chi_Delta_t}).
\\
ii) $\mu\neq0$ and $\Delta/t=1$

There are two poles $x=-2t/\mu$, $-\mu/2t$
and $\chi$ is given by Eq.~(\ref{eq:chi_Delta_t}).
\\
iii) In other cases ($\mu\neq0$ and $\Delta/t\neq1$)

There are four poles.
\begin{align}
   x
   =&
   \frac{%
      -\mu/2\pm\sqrt{\mu^2/4-t^2+\Delta^2}
   }%
   {%
      t\pm\Delta
   }.
\end{align}
Note that poles
\begin{align}
   x_{++}
   =&
   \frac{%
      -\mu/2+\sqrt{\mu^2/4-t^2+\Delta^2}
   }%
   {%
      t+\Delta
   },
   \\
   x_{--}
   =&
   \frac{%
      -\mu/2-\sqrt{\mu^2/4-t^2+\Delta^2}
   }%
   {%
      t-\Delta
   },
   \\
   x_{+-}
   =&
   \frac{%
      -\mu/2+\sqrt{\mu^2/4-t^2+\Delta^2}
   }%
   {%
      t-\Delta
   },
   \\
   x_{-+}
   =&
   \frac{%
      -\mu/2-\sqrt{\mu^2/4-t^2+\Delta^2}
   }%
   {%
      t+\Delta
   },
\end{align}
satisfy
\begin{align}
   x_{++}x_{--}=x_{+-}x_{-+}=1.
\end{align}
Then there are two poles in the unit circle and we have to check which are
located in the circle.

Note that
$|x_{++}|<|x_{--}|$ 
is satisfied for all parameter range.

\subsubsection{$\mu^2/4-t^2+\Delta^2<0$}
In this case 
$|x_{+-}|>|x_{-+}|$ and $\mu<2t$ (QCR1 or QCR2) are satisfied and $\chi$ is
\begin{align}
   \chi
   =&
   \chi_1
   =
   \frac{2t^2}{(\mu^2-4t^2)\Delta^2}
   -2
   \frac{\mu^2+4t^2}{(\mu^2-4t^2)^2}.
\end{align}
For QCR1 ($|\Delta/(t-\mu/2)|\gg 1$), 
$\chi$ is given by
\begin{align}
   \chi
   \sim
   -2
   (\mu^2+4t^2)
   \frac{1}{(\mu+2t)^2}
   \frac{1}{(\mu-2t)^2},
\end{align}
and for QCR2 ($\Delta^2/(t^2-\mu^2/4)\ll 1$), $\chi$ is given by
\begin{align}
   \chi
   \sim
   \frac{2t^2}{\Delta^2}
   \frac{1}{\mu+2t}
   \frac{1}{\mu-2t}.
\end{align}

\subsubsection{$\mu^2/4-t^2+\Delta^2>0$}
There are two cases.
\\
1)$\mu<2t$ (QCR1 or QCR2)
\\
QCR1 ($\Delta^2/(4t^2-\mu^2)\gg 1$) is included in this case.

In this case, the relation
$|x_{+-}|>|x_{-+}|$ holds and $\chi$ is
\begin{align}
   \chi
   =&
   \chi_1
   \sim
   -2
   (\mu^2+4t^2)
   \frac{1}{(\mu+2t)^2}
   \frac{1}{(\mu-2t)^2},
\end{align}
\\
2)$\mu>2t$ (QCR1 or QCR3)

In this case, the relation
$|x_{+-}|<|x_{-+}|$ holds and $\chi$ is
\begin{align}
   \chi
   =&
   \chi_2
   =
   \frac{\Delta \mu t/2}
   {(\mu^2/4-t^2)^2\sqrt{\Delta^2+\mu^2/4-t^2}}.
\end{align}

In the case $\Delta^2/(t^2-\mu^2/4)\gg 1$ (QCR1), $\chi_2$ can be
expanded as
\begin{align}
   \chi_2
   \sim
   -
   \frac{1}{(\mu-2t)^2}
   +
   \mathcal{O}((\mu-2t)^{-1}).
\end{align}

In the case $|\Delta/(t-\mu/2)|\ll 1$ (QCR3), $\chi_2$ can be
expanded as
\begin{align}
   \chi_2
   \sim
   \frac{16\mu t\Delta}{(\mu^2-4t^2)^{5/2}}
   -
   \frac{32\mu t \Delta^3}{(\mu^2-4t^2)^{7/2}}
   +
   \mathcal{O}(\Delta^5).
\end{align}

\subsection{Critical exponent of $\chi$ for Kitaev chain\label{sec:chi_app}}

The low energy action for the Kitaev chain is given by
\begin{align}
   S
   =
   \frac{1}{2}
   \int\mathrm{d}\tau\mathrm{d}x
   \Psi^\dag
   \left[
      \partial_\tau
      \tau_0
      +
      v
      \partial_x
      \tau_2
      +
      \left(
         m
         -
         \Lambda
         \partial_x^2
      \right)
      \tau_3
   \right]
   \Psi,
   \label{eq:action_sup}
\end{align}
with $\Psi=(\psi,\psi^\dag)^\mathrm{T}$.
For the Kitaev chain, $v$, $m$ and $\Lambda$ are given by
$v=\Delta$, $m=\delta \mu=\mu-\mu_\mathrm{c}$ and $\Lambda=t$, respectively, and
this action describes the low-energy excitations that appear around wavevector $\pi$.

Let us consider the following scaling transformation:
\begin{align}
   x'
   =&
   x
   e^{-l},
   \\
   \tau'
   =&
   \tau
   e^{\dim[\tau]l},
   \label{eq:scale_z}
   \\
   \psi'
   =&
   \psi
   e^{\dim[\psi]l},
   \\
   v'
   =&
   v
   e^{\dim[v]l},
   \label{eq:scale_v}
   \\
   m'
   =&
   m
   e^{\dim[m]l},
   \label{eq:scale_m}
\end{align}
The conditions for a scale-invariant action are $\dim \tau = -2$, $\dim \psi = 1/2$, $\dim v = 1$, and $\dim m=2$.
Since $w_\mathrm{bulk}(z)$ is an even function of $z$, we can expand it as
\begin{align}
   w_\mathrm{bulk}(z)
   =&
   \sum_{n=0}^\infty
   a_{2n}z^{2n},
\end{align}
where $W = a_0$ and $\chi = a_2$.
$w_\mathrm{bulk}(z)$ is dimensionless and therefore its scaling dimension is zero.
As explained in the main text, $a_{2n}$ is expressed near the critical points as
\begin{align}
   a_{2n}
   \sim
   v^{\alpha_n}
   m^{\beta_n}.
\end{align}
By applying scale transformation in Eqs.~(\ref{eq:scale_z}), (\ref{eq:scale_v}) 
and (\ref{eq:scale_m}), we get
\begin{align}
   \al_n \dim [v] + \beta_n \dim [m] + 2 n \dim [z] = 0.
   \label{eq:relation}
\end{align}
Here we have used the relation $\dim z = - \dim \tau$.
In the general action in Eq.\eqref{eq:action_sup}, the relation between $\al_n$ and $\beta_n$ 
is determined by Eq.~\eqref{eq:relation}, but we need more information to determine the concrete value of exponents.

To specify the functional form in more details, 
we consider the three quantum critical regions (QCR): 
$|\Delta| \gg |\delta \mu|$ (QCR1), 
$|\Delta| \ll |\delta \mu| $, $\delta \mu<0$ (QCR2) 
and $|\Delta| \ll |\delta \mu| $, $\delta \mu>0$ (QCR3).
See also Fig.~4 in the main text.
For QCR1, the criticality is determined by the Ising universality class, 
and the $\partial_x^2$ term in Eq.~\eqref{eq:action_sup} can be neglected in the low-energy region.
In this case, the effective action is 
\begin{align}
   \tilde S 
   = 
   \frac 1 2 \int \diff \tilde \tau \diff \tilde x 
   \tilde{\Psi}^\dg 
   \left[
      \partial_{\tilde \tau} \tau_0 + \tilde v \tau_2 + \tilde m\tau_3
   \right]
   \tilde \Psi
   \label{eq:action_tilde}
\end{align}
where $\tilde v = v$ and $\tilde m = m$.
The action is invariant
if $\dim [\tilde \tau]=-1$, $\dim [\tilde v]=0$ and $\dim [\tilde m]=1$.
In this regime, we can write the coefficient as $a_{2n} \sim \tilde m^{\beta'_n}$.
Then we can get $\beta'_n = -2n$ from the scaling invariance of $w_{\rm bulk}$.
Correspondingly, we get $\beta_n = -2n$, and from Eq.(\ref{eq:relation}) we get $\al_n = 0$.
We find the odd-frequency pair amplitude with the form 
\begin{align}
   F_\mathrm{edge}^\mathrm{odd}(z) 
   = 
   z^{-1}\mathcal W_1
   \left(
      \frac{z^2}{\delta\mu^2}
   \right),
\end{align}
indicating 
\begin{align}
   \chi \sim \delta\mu^{-2}.
\end{align}

For QCR2, we focus on the criticality that appears when $\Delta \to 0$, 
and then we get the behavior with respect to $\delta \mu$ by utilizing Eq.~\eqref{eq:relation}.
The position of low-energy excitation shifts from $k=0$ to $k=2\sqrt{|m|/\Lambda}$.
The corresponding low-energy action is given by Eq.~\eqref{eq:action_tilde}
with the velocity $\tilde v \sim \sqrt{\Lambda m}$ and mass $\tilde m \sim v\sqrt{m/\Lambda}$.
The scaling invariance requires
$\dim [\tilde \tau]=-1$, $\dim [\tilde v]=0$ and $\dim [\tilde m]=1$.
In this regime, we can write the coefficient as $a_{2n} \sim \tilde m^{\beta''_n}$.
Then we can get $\beta''_n = -2n$ from the scaling invariance of $w_{\rm bulk}$, and we then get $\beta_n = -n$.
From Eq.~(\ref{eq:relation}) we also get $\al_n = -2n$.
We find the odd-frequency pair amplitude with the form 
\begin{align}
   F_\mathrm{edge}^\mathrm{odd}(z) 
   = 
   z^{-1} \mathcal W_2 
   \left(
      \frac{tz^2}{\delta \mu \Delta^2}
   \right),
\end{align}
and 
\begin{align}
   \chi \sim \Delta^{-2}\delta \mu^{-1}.
\end{align}

Finally we consider QCR3.
In this case, 
we cannot construct another effective action like Eq.~(\ref{eq:action_tilde}).
The simplest way to determine the critical behavior of $F_\mathrm{edge}^\mathrm{odd}(z)$ 
is to consider the power-counting of $\Delta$ in the concrete expression of $w_\mathrm{bulk}(z)$ in Eq.~\eqref{eq:w_bulk_Kitaev_gen}.
Taking $\Delta \to 0$, one can easily see $w_\mathrm{bulk}(z) \sim \Delta$, 
meaning $\al_n=1$ and then $\beta_n = -\tfrac 1 2 - 2n$.
We get 
\begin{align}
   F_\mathrm{edge}^\mathrm{odd}(z) 
   = 
   z^{-1} \Delta (t\delta \mu)^{-1/2} \mathcal W_{3}
   \left(
      \frac{z^2}{\delta \mu^2}
   \right),
\end{align}
and 
\begin{align}
   \chi\sim \Delta \delta \mu^{-5/2}.
\end{align}
Thus the numerical results are completely interpreted in terms of the generalized scaling theory.
\section{Nanowire with Rashba spin-orbit interaction and Zeeman field\label{sec:SC_app1}}
\subsection{SBBC of nanowire\label{sec:SC_app1_nano_sbbc}}
\begin{figure*}[htbp]
   \centering
   \includegraphics[width = 17cm]{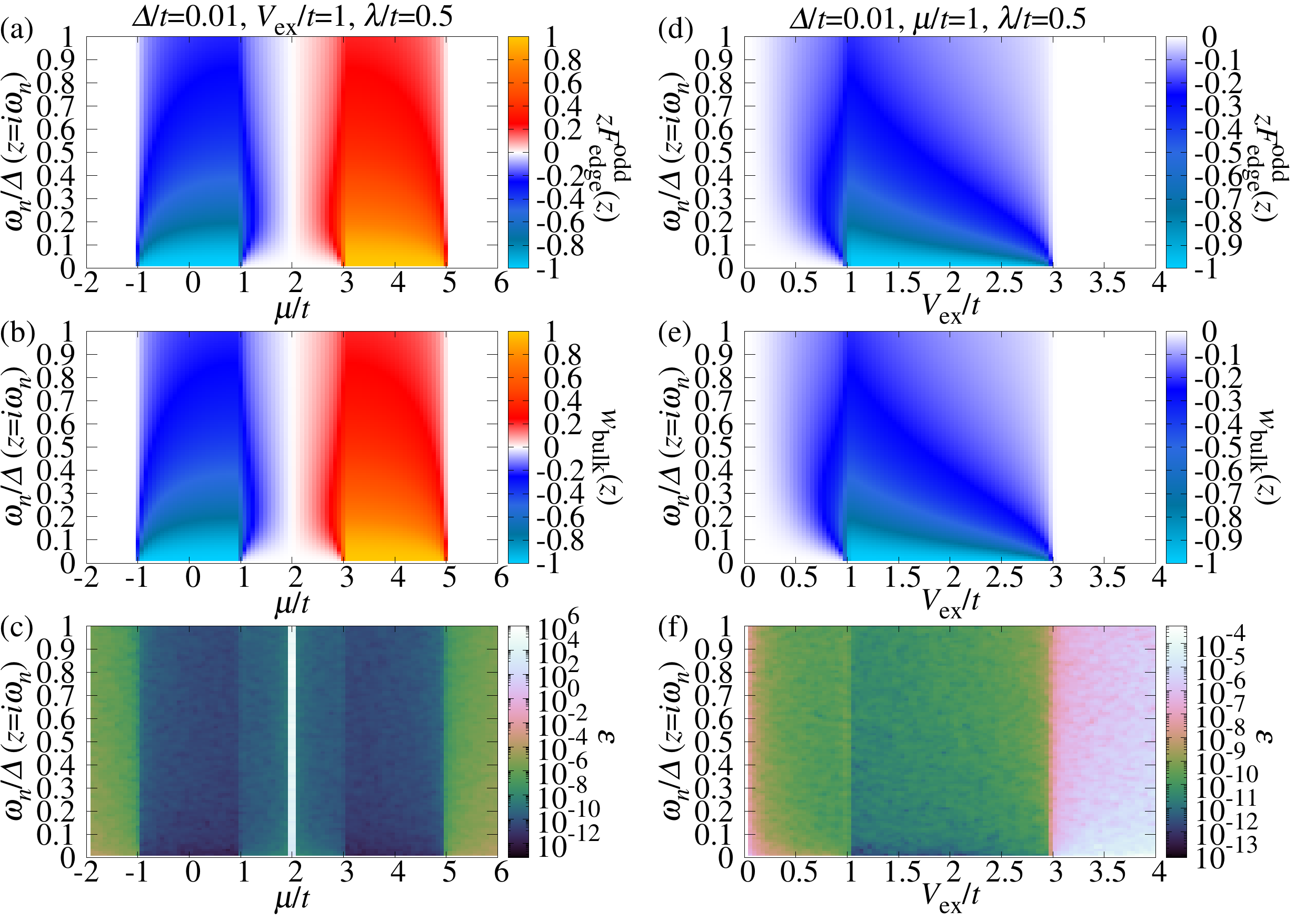}
   \caption{%
      (a) The real part of $zF_\mathrm{edge}^\mathrm{odd}(z)$,
      (b) the real part of $w_\mathrm{bulk}(z)$, and 
      (c) $\varepsilon$ are plotted 
      as functions of $\mu/t$ and $\omega/\Delta$ for
      $\Delta/t=0.01$, $V_\mathrm{ex}/t=1$, and $\lambda/t=0.5$.
      (d) The real part of $zF_\mathrm{edge}^\mathrm{odd}(z)$,
      (e) the real part of $w_\mathrm{bulk}(z)$, and
      (f) $\varepsilon$ are plotted 
      as functions of $V_\mathrm{ex}/t$ and $\omega/\Delta$ for
      $\Delta/t=0.01$, $\mu/t=1$ and, $\lambda/t=0.5$.
      Note that $zF_\mathrm{edge}^\mathrm{odd}(z)$ and $w_\mathrm{bulk}(z)$ are real valued function
      for purely imaginary $z$.
   }%
   \label{fig:nano_wire_3d}
\end{figure*}

In this section, we numerically calculate $F_\mathrm{edge}^\mathrm{odd}(z)$ and $w_\mathrm{bulk}(z)/z$
for the system which is the one-dimensional $s$-wave superconductor with Rashba spin-orbit coupling
and Zeeman field.
The Kitaev chain does not have a spin degree of freedom but this model has it and is more complicated system.
At first, we show $zF_\mathrm{edge}^\mathrm{odd}(z)$, $w_\mathrm{bulk}(z)$ and $\varepsilon$ for the nanowire
in Fig.~\ref{fig:nano_wire_3d}.
When we calculate $\mathrm{Tr}_j$ in $F_\mathrm{edge}^\mathrm{odd}(z)$, we take $10^5$ sites from the surface and
when we calculate integral in $w_\mathrm{bulk}(z)$, we also take $10^5$ wave numbers.
In Figs.~\ref{fig:nano_wire_3d}(c) and (f), $\varepsilon$ becomes large where 
the value of $F_\mathrm{edge}^\mathrm{odd}(z)$ and $w_\mathrm{odd}(z)$ are approximately zero
[see Figs.~\ref{fig:nano_wire_3d}(a), (b), (d) and (e)].
In Figs.~\ref{fig:nano_wire_3d}(c), at $\mu/t=2$, $\varepsilon$ is very large because
the order of $zF_\mathrm{edge}^\mathrm{odd}(z)$ is $10^{-11}$ and that of $w_\mathrm{bulk}(z)$
is $10^{-15}$ and they are almost zero.

\subsection{Criticality of nanowire\label{sec:SC_app1_crit}}
The low energy effective action is given by Eq.~(\ref{eq:action_sup}).
Let $\delta\mu_{1,3}$ and $\delta V_{1,2}$ be
\begin{align}
   \delta\mu_{1}
   =&
   \mu-\mu_{\mathrm{c}1},
   \\
   \delta\mu_{3}
   =&
   \mu-\mu_{\mathrm{c}3},
   \\
   \delta V_1
   =&
   V_\mathrm{ex}-V_{\mathrm{c}1},
   \\
   \delta V_2
   =&
   V_\mathrm{ex}-V_{\mathrm{c}2},
\end{align}
with $\mu_{\mathrm{c}1}=-\sqrt{V_\mathrm{ex}^2-\Delta^2}$,
$\mu_{\mathrm{c}3}=4t-\sqrt{V_\mathrm{ex}^2-\Delta^2}$,
$V_{\mathrm{c}1}=\sqrt{\mu^2+\Delta^2}$, and
$V_{\mathrm{c}2}=\sqrt{(4t-\mu)^2+\Delta^2}$.

We only show the results near the topological quantum transition points
at $\mu=\mu_{\mathrm{c}1}$, $\mu_{\mathrm{c}3}$.
The critical behavior near
$
\mu_{\mathrm{c}2}=\sqrt{V_\mathrm{ex}^2-\Delta^2}$ and
$\mu_{\mathrm{c}4}=4t+\sqrt{V_\mathrm{ex}^2-\Delta^2}$
are the same as that near
$\mu_{\mathrm{c}1}$ and
$\mu_{\mathrm{c}3}$,
respectively.
We suppose $|\Delta|\ll t$ in the following discussion.

\subsubsection{Near the transition point $V_\mathrm{ex}=V_{\mathrm{c}1}$}


(i) $|\delta\mu_1|\ll t$.

The constants $v$, $m$ and $\Lambda$ in Eq.~(\ref{eq:action_sup}) are
given by
\begin{align}
   v
   \sim&
   \frac{\lambda\Delta}{V_\mathrm{ex}},
   \label{eq:delmu1_1}
   \\
   m
   \sim&
   \delta\mu_1,
   \label{eq:delmu1_2}
   \\
   \Lambda
   \sim&
   \frac{\lambda^2}{2|\mathrm{V_\mathrm{ex}}|}
   +
   t.
   \label{eq:delmu1_3}
\end{align}

(ii) $|\delta V_1|\ll t$.

$v$, $m$ and $\Lambda$ in Eq.~(\ref{eq:action_sup}) are
given by
\begin{align}
   v
   \sim&
   \frac{\lambda\Delta}{\mu},
   \label{eq:delV1_1}
   \\
   m
   \sim&
   \delta V_1,
   \label{eq:delV1_2}
   \\
   \Lambda
   \sim&
   \frac{\lambda^2+2t\mu}{2|\mu|}.
   \label{eq:delV1_3}
\end{align}

Note that the roles of the $V_\mathrm{ex}$ and $\mu$ in the cases (i) and (ii) are exchanged.
With these coefficients, the critical behaviors can be explained based on the scaling theory described in the previous section.

\subsubsection{Near the transition point $V_\mathrm{ex}=V_{\mathrm{c}2}$}

(i) $|\delta\mu_3|\ll t$.

$v$, $m$ and $\Lambda$ in Eq.~(\ref{eq:action_sup}) are
given by
\begin{align}
   v
   \sim&
   \frac{\lambda\Delta}{V_\mathrm{ex}},
   \label{eq:delmu2_1}
   \\
   m
   \sim&
   \delta \mu_3,
   \label{eq:delmu2_2}
   \\
   \Lambda
   \sim&
   \frac{\lambda^2}{2|V_\mathrm{ex}|}+t.
   \label{eq:delmu2_3}
\end{align}

(ii) $|\delta V_2|\ll t$.

$v$, $m$ and $\Lambda$ in Eq.~(\ref{eq:action_sup}) are
given by
\begin{align}
   v
   \sim&
   \frac{\lambda\Delta}{\mu-4t},
   \label{eq:delV2_1}
   \\
   m
   \sim&
   \delta V_2,
   \label{eq:delV2_2}
   \\
   \Lambda
   \sim&
   \frac{\lambda^2-2t(\mu-4t)}{2|\mu-4t|}.
   \label{eq:delV2_3}
\end{align}

\subsubsection{Numerical results}

\begin{figure*}[htbp]
   \centering
   \includegraphics[width = 17cm]{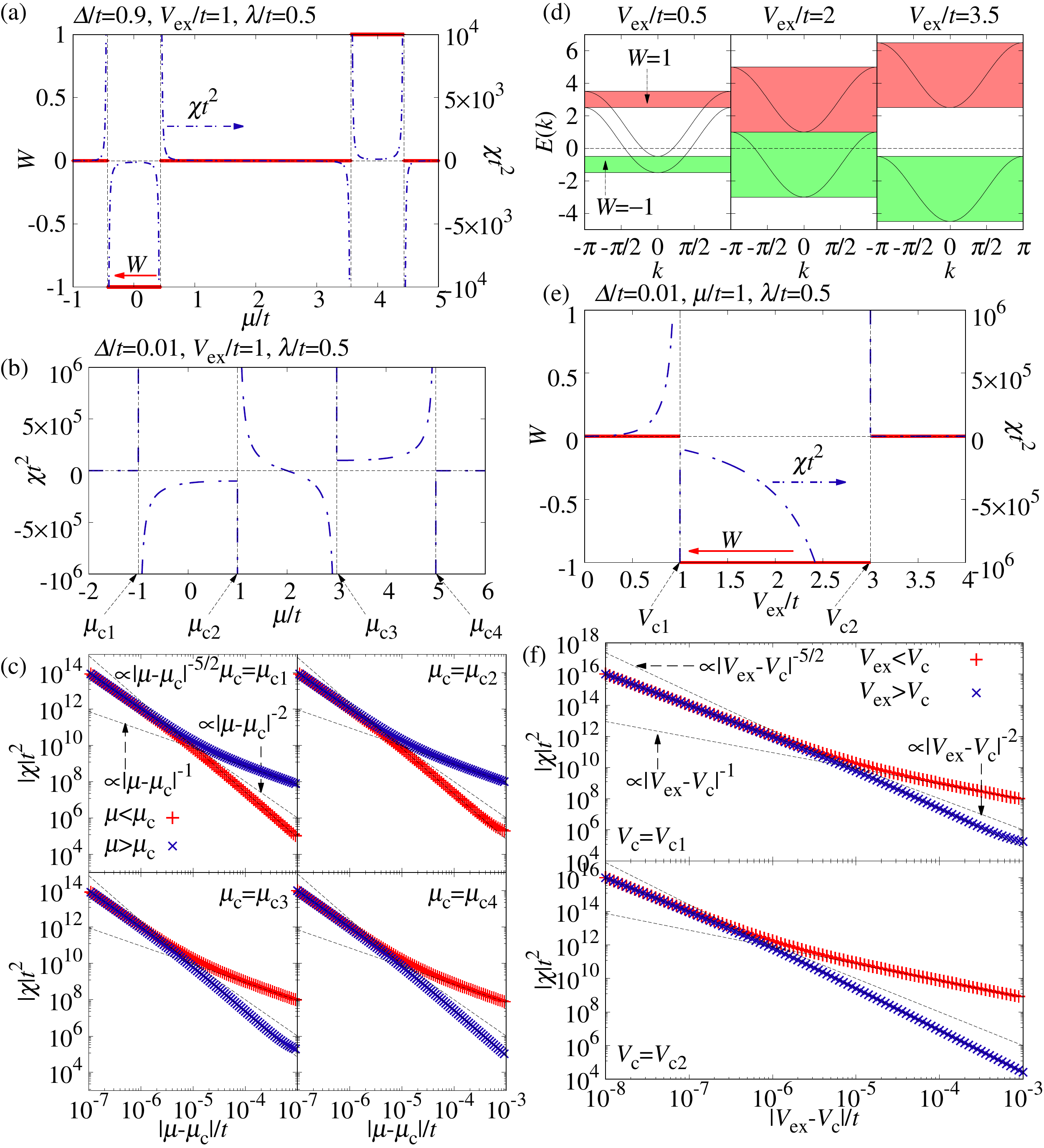}
   \caption{%
      (a) $W$ (left vertical axis) and $\chi t^2$ (right vertical axis) are plotted as
      a function of $\mu/t$ for $\Delta/t=0.9$, $V_\mathrm{ex}/t=1$ and $\lambda/t=0.5$.
      (b) $\chi t^2$ is plotted as a function of $\mu/t$ for $\Delta/t=0.01$, $V_\mathrm{ex}/t=1$ and $\lambda/t=0.5$
      with $\mu_{\mathrm{c}1}=-\sqrt{V_\mathrm{ex}^2-\Delta^2}$, 
      $\mu_{\mathrm{c}2}=\sqrt{V_\mathrm{ex}^2-\Delta^2}$,
      $\mu_{\mathrm{c}3}=4t-\sqrt{V_\mathrm{ex}^2-\Delta^2}$, 
      and $\mu_{\mathrm{c}4}=4t+\sqrt{V_\mathrm{ex}^2-\Delta^2}$.
      (c) $|\chi|t^2$ is plotted as a function of $|\mu-\mu_\mathrm{c}|/t$ close to 
      $\mu=\mu_{\mathrm{c}1}$, $\mu=\mu_{\mathrm{c}2}$, $\mu=\mu_{\mathrm{c}3}$, 
      and $\mu=\mu_{\mathrm{c}4}$.
      (d) Energy dispersion is plotted as a function of $k$ with $V_\mathrm{ex}=0.5$, $V_\mathrm{ex}=2$
      and $V_\mathrm{ex}=3.5$ for $\Delta/t=0.01$, $\mu/t=1$ and $\lambda/t=0.5$.
      Green shaded area and red shaded one is a topological regime with $W=-1$ for green and $W=1$ for red, respectively.
      (e) $W$ (left vertical axis) and $\chi t^2$ (right vertical axis) are plotted 
      as a function of $V_\mathrm{ex}/t$ for $\Delta/t=0.01$, $\mu/t=1$ and $\lambda/t=0.5$
      with $V_{\mathrm{c}1}=\sqrt{\mu^2-\Delta^2}$, $V_{\mathrm{c}2}=\sqrt{{(4t-\mu)}^2-\Delta^2}$.
      $W$ and $\chi$ are even function of $V_\mathrm{ex}$.
      (f) $|\chi|t^2$ is plotted as a function of $|V_\mathrm{ex}-V_\mathrm{c}|/t$ close to 
      $V_\mathrm{ex}=V_{\mathrm{c}1}$ and $V_\mathrm{ex}=V_{\mathrm{c}2}$.
   }%
   \label{fig:nano_wire_slope_mu}
\end{figure*}

Then we discuss about $\chi$.
In Fig.~\ref{fig:nano_wire_slope_mu}(a), $W$ and $\chi$ are plotted as a function of $\mu/t$
for $\Delta/t=0.9$, $V_\mathrm{ex}/t=1$ and $\lambda/t=0.5$.
As mentioned in the main text, $\chi$ diverges slower than the case with $\Delta/t=0.01$.
In Fig.~\ref{fig:nano_wire_slope_mu}(b), $\chi/t$ is plotted as a function of $\mu/t$
and there are four topological transition points 
$\mu_{\mathrm{c}1}$ to $\mu_{\mathrm{c}4}$.
We show $|\chi|$ near topological transition points in Fig.~\ref{fig:nano_wire_slope_mu}(c).
The power of divergence is $|\mu-\mu_\mathrm{c}|^{-2}$ very close to the
transition point for all topological transition points.
As explained in the main text, the divergence far from transition points
are $\delta\mu^{-1}$ or $\delta\mu^{-5/2}$ with $\delta\mu=\mu-\mu_\mathrm{c}$.

Next, we discuss about $V_\mathrm{ex}$ dependence of $\chi$.
In Fig.~\ref{fig:nano_wire_slope_mu}(d) energy dispersion is plotted for 
$V_\mathrm{ex}/t=0.5$ (non-topological), $V_\mathrm{ex}/t=2$ (topological) and 
$V_\mathrm{ex}/t=3.5$ (non-topological) with $\Delta/t=0.01$, $\mu/t=1$ and $\lambda/t=0.5$.
$W$ and $\chi$ are plotted as a function of $V_\mathrm{ex}$ in Fig.~\ref{fig:nano_wire_slope_mu}(e).
There are two topological transition points 
$V_{\mathrm{c}1}$ and $V_{\mathrm{c}2}$.
$|\chi|$ is plotted near the transition points shown in Fig.~\ref{fig:nano_wire_slope_mu}(f).
The power of divergence is also $|V_\mathrm{ex}-V_\mathrm{c}|^{-2}$
very close to the transition point for both transition points.
The divergence behavior changes if we look at far from transition point.
The power of divergence is $\delta V^{-1}$ or $\delta V^{-5/2}$ 
with $\delta V=V-V_\mathrm{c}$.

\section{$d_{x^2-y^2}$-wave SC on square lattice\label{sec:SC_app3}}
\subsection{SBBC of $d_{x^2-y^2}$-wave SC\label{sec:SC_app3_sbbc}}
In this section we calculate $F_\mathrm{edge}^\mathrm{odd}(z)$ and $w_\mathrm{bulk}(z)/z$ for 
$d_{x^2-y^2}$-wave superconductor on a square lattice with the (11) surface.

The Hamiltonian with the periodic boundary condition is
\begin{align}
   {\cal H}
   =&
   \sum_\mathbf{k}
   C_\mathbf{k}^\dag
   H(\mathbf{k})
   C_\mathbf{k},
   \\
   H(\mathbf{k})
   =&
   \left[
      -
      \mu\eta_0
      +
      \varepsilon_1(\mathbf{k})
      \eta_1
      +
      \varepsilon_2(\mathbf{k})
      \eta_2
   \right]
   \sigma_0
   \tau_3
   \nonumber\\
   &
   +
   \left[
      \Delta_1
      \left(
         \mathbf{k}
      \right)
      \eta_1
      +
      \Delta_2
      \left(
         \mathbf{k}
      \right)
      \eta_2
   \right]
   i\sigma_2
   i
   \tau_2,
\end{align}
with
\begin{align}
   \varepsilon_1(\mathbf{k})
   =&
   -t
   \left[
      1+\cos k_x + \cos k_y + \cos(k_x-k_y)
   \right],
   \\
   \varepsilon_2(\mathbf{k})
   =&
   t
   \left[
      \sin k_x + \sin k_y + \sin(k_x-k_y)
   \right],
   \\
   \Delta_1(\mathbf{k})
   =&
   -
   \Delta
   \left[
      1-\cos k_x - \cos k_y + \cos(k_x-k_y)
   \right],
   \\
   \Delta_2(\mathbf{k})
   =&
   \Delta
   \left[
      \sin k_x + \sin k_y - \sin(k_x-k_y)
   \right],
   \\
   C_\mathbf{k}
   =&
   \left(
      c_{A,\mathbf{k},\uparrow}\:
      c_{B,\mathbf{k},\uparrow}\:
      c_{A,\mathbf{k},\downarrow}\:
      c_{B,\mathbf{k},\downarrow}\:
   \right.
   \nonumber
   \\
   &
   \left.
      \hspace{5mm}
      c_{A,-\mathbf{k},\uparrow}^\dag\:
      c_{B,-\mathbf{k},\uparrow}^\dag\:
      c_{A,-\mathbf{k},\downarrow}^\dag\:
      c_{B,-\mathbf{k},\downarrow}^\dag
   \right)^\mathrm{T},
\end{align}
where $\eta_0$ and $\sigma_0$ are identity matrices in sublattice space, and spin 
space, respectively.
$\eta_\mu$, $\sigma_\mu$ and $\tau_\mu$ ($\mu=1,2,3$)
are Pauli matrices in sublattice, spin, and particle-hole space, respectively.
The unit cell used in this Hamiltonian is shown in Fig.~\ref{fig:pic_dx2y2} (a)
The chiral operator for the bulk (periodic) system is
\begin{align}
   \Gamma 
   =&
   \eta_0\sigma_2\tau_1.
\end{align}

The Hamiltonian with the (11) surface for the semi-infinite system is
\begin{align}
   {\cal H}
   =&
   C^\dag
   \begin{pmatrix}
      \ddots &\ddots
      \\
      \ddots&\hat{u} &\hat{t}
      \\
      &\hat{t}^\dag & \hat{u} &\hat{t}
      \\
      &&\hat{t}^\dag & \hat{u}
   \end{pmatrix}
   C,
   \\
   \hat{u}
   =&
   \begin{pmatrix}
      -\mu/2 &0&0&0
      \\
      0&-\mu/2 &0&0
      \\
      0&0&\mu/2 &0
      \\
      0&0&0&\mu/2 
   \end{pmatrix},
   \\
   \hat{t}
   =&
   \begin{pmatrix}
      -t\cos \frac{k_y}{2} &0&0&i\Delta\sin\frac{k_y}{2}
      \\
      0&-t\cos \frac{k_y}{2} &-i\Delta\sin \frac{k_y}{2}&0
      \\
      0&-i\Delta\sin \frac{k_y}{2}&t\cos \frac{k_y}{2} &0
      \\
      i\Delta\sin \frac{k_y}{2}&0&0&t\cos \frac{k_y}{2}
   \end{pmatrix},
\end{align}
with
\begin{align}
   C
   =&
   \left(
      \ldots,
      c_{1,k_y,\uparrow},
      c_{1,k_y,\downarrow},
      c_{1,-k_y,\uparrow}^\dag,
      c_{1,-k_y,\downarrow}^\dag
   \right)^\mathrm{T}.
\end{align}
The chiral operator for the semi-infinite system is
\begin{align}
   \Gamma
   =&
   \mathrm{diag}(\ldots,\sigma_2\tau_1,\sigma_2\tau_1).
\end{align}

In Fig.~\ref{fig:pic_dx2y2}(b), we show $F_\mathrm{edge}^\mathrm{odd}(z)$ 
for $\Delta/t=0.1$ and $\mu/t=-1$ 
and it is odd function of $k_y$.
When we calculate $\mathrm{Tr}_j$ in $F_\mathrm{edge}^\mathrm{odd}(z)$, 
we take $10^4$ sites from the surface and when we calculate integral in 
$w_\mathrm{bulk}(z)$, we also take $10^4$ wave numbers.
$\varepsilon$ is plotted as functions of $k_y$ and $\omega_n/\Delta$ in Fig.~\ref{fig:pic_dx2y2}(c) 
and is less than $10^{-7}$.
At $k_y=0$, there is no value because $F_\mathrm{edge}^\mathrm{odd}(z)$ and 
$w_\mathrm{bulk}(z)$ are exactly zero.
In Fig.~\ref{fig:pic_dx2y2}(d), we show $W$ and $\chi$ as a function of $k_y$
and there are three topological transition points 0 and $\pm2\arccos(-\mu/4t)$.
\subsection{Criticality of $d_{x^2-y^2}$-wave SC\label{sec:SC_app3_crit}}
There are three topological transition points:
$k_y=0$ and $k_y=\pm k_\mathrm{c}$ with $k_\mathrm{c}=2\arccos(-\mu/4t)$.

\subsubsection{Criticality at $k_y = \pm k_c$}
The effective low energy action near $k_y=\pm k_\mathrm{c}$ is given by 
Eq.~(\ref{eq:action_sup})
with
\begin{align}
   v
   \sim&
   \Delta
   \sqrt{4-\frac{\mu^2}{4t^2}},
   \\
   m
   \sim&
   t^2
   \delta k
   \sqrt{%
      4
      -
      \frac{\mu^2}{4t^2}
   },
   \\
   \Lambda
   \sim&
   \frac{\mu}{8},
\end{align}
with $\delta k=k_x-k_\mathrm{c}$.
The scaling behavior can be understood based on Eq.~(\ref{eq:action_sup}) as in the previous sections.

\subsubsection{Criticality at $k_y = 0$}

The effective low energy action near $k_y=0$ is given by
\begin{align}
   S
   =&
   \frac{1}{2}
   \int\mathrm{d}\tau\mathrm{d}x
   \Psi^\dag
   \left[
      \partial_\tau
      \tau_0
      +
         \tilde{v}
         \partial_x
      \tau_3
      +
      \tilde{m}
      \tau_1
   \right]
   \Psi,
\end{align}
with
\begin{align}
   \tilde{v}
   \sim&
   2t\sin\left(\frac{k_\mathrm{c}}{2}\right),
   \\
   \tilde{m}
   \sim&
   2k_y\Delta
   \sin
   \left(
      \frac{k_\mathrm{c}}{2}
   \right).
\end{align}
Then by using scale transformation, we get
\begin{align}
   \dim[\tau]
   =&
   -1,
   \\
   \dim[\psi]
   =&
   1/2,
   \\
   \dim[k_y]
   =&
   1/2,
   \\
   \dim[\Delta]
   =&
   1/2.
\end{align}
Here we use the fact that $k_y$ and $\Delta$ are symmetric in the effective
action.
Then the generalized winding number can be expanded as
\begin{align}
   w_\mathrm{bulk}(z)
   \sim
   \sum_n
   \tilde{a}_{2n}
   {(k_y\Delta)}^{\gamma_n}z^{2n}.
\end{align}
By using scale transformation, we obtain
\begin{align}
   &
   2n\dim[z]
   +
   \gamma_n
   \left(
      \dim[k_y]+\dim[\Delta]
   \right)
   =0,
   \nonumber\\
   \Leftrightarrow
   &
   2n+\gamma_n=0.
   \label{eq:rel_crit_d}
\end{align}

%

Then the generalized winding number
is obtained as
\begin{align}
   w_\mathrm{bulk}(z)
   \sim&
   \sum_n
   \tilde{a}_{2n}
   (k_y\Delta)^{-2n}z^{2n},
\end{align}
and $\chi$ behaves as
\begin{align}
   \chi
   \sim&
   (k_y\Delta)^{-2}.
\end{align}

\begin{figure*}[htbp]
   \centering
   \includegraphics[width = 17cm]{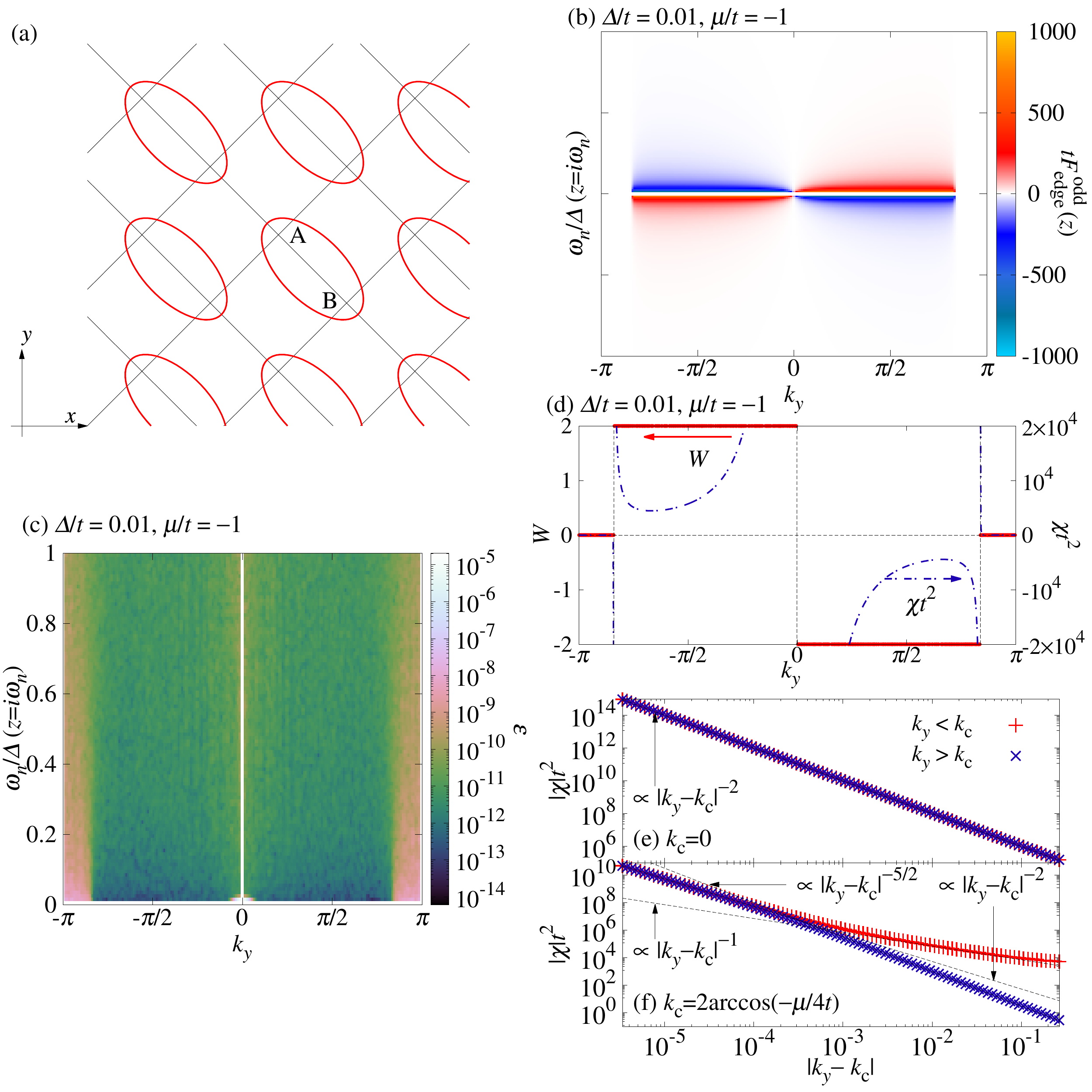}
   \caption{%
      (a)Schematic illustration of the unit cell.  A and B indicate sublattices.
      (b) Imaginary part of $F_\mathrm{edge}^\mathrm{odd}(z)$ is plotted as functions of
      $k_y$ and $\omega_n/\Delta$ with $\Delta/t=0.01$ and $\mu/t=-1$.
      (c) $\varepsilon$ is plotted as functions of $k_y$ and $\omega_n/\Delta$ with $\Delta/t=0.01$ and $\mu/t=-1$.
      At $k_y=0$, there is no value because $w_\mathrm{bulk}(z)$ and $F_\mathrm{edge}^\mathrm{odd}(z)$ are exactly zero.
      (d) $\chi t^2$ is plotted as a function of $k_y$.
      $|\chi| t^2$ is plotted as a function of $|k_y-k_\mathrm{c}|$ near 
      (e) $k_\mathrm{c}=0$ and (f) $k_\mathrm{c}=2\arccos(-\mu/4t)$.
      Dotted lines in (b) and (c) are proportional to $|k_y-k_\mathrm{c}|^{-2}$.
   }%
   \label{fig:pic_dx2y2}
\end{figure*}

$|\chi|$ is shown in Figs.~\ref{fig:pic_dx2y2}(e) near $k_y=0$ and (f) near $k_y=k_\mathrm{c}$.
In both cases, close to the quantum transition point, 
the power of the divergence is $k_y^{-2}$ and $|k_y-k_\mathrm{c}|^{-2}$, respectively.
Away from the transition point $k_y=k_\mathrm{c}$, the power of divergence 
is $(k_y-k_\mathrm{c})^{-1}$ [$(k_y-k_\mathrm{c})^{-5/2}$] for
$k_y<k_\mathrm{c}$ ($k_y>k_\mathrm{c}$).



\end{document}